\documentclass[a4paper,11pt]{article}
\usepackage{jcappub}
\usepackage{amsmath}
\usepackage{graphicx}
\usepackage{latexsym}
\usepackage{xspace}
\usepackage{color}
\usepackage{bm}
\usepackage{relsize}
\usepackage{tabularx}
\usepackage{multirow}
\usepackage{amssymb}
\usepackage[table]{xcolor}
\usepackage{braket}
\usepackage{comment}
\usepackage[utf8]{inputenc}
\usepackage{slashed}
\usepackage{empheq}
\usepackage{subfigure}
\usepackage{here}

\makeatletter
\gdef\@fpheader{}
\g@addto@macro\bfseries{\boldmath}
\makeatother

\newcommand{\erfc}{{\mathrm{erfc}}}

\newcommand{\Pfpt}{P_{\sss{\mathrm{FPT}}}}
\newcommand{\sigmaL}{{\sigma_{\mathrm{L}}}}
\newcommand{\sigmaR}{{\sigma_{\mathrm{R}}}}

\newcommand{\ie}{\textsl{i.e.~}}

\newcommand{\dd}{\mathrm{d}}

\newcommand{\sss}[1]{{\scriptscriptstyle{#1}}}
\newcommand{\boldmathsymbol}[1]{{\ensuremath{\boldsymbol{#1}}}}

\newcommand{\uPl}{\mathrm{Pl}}

\newcommand{\uend}{\mathrm{end}}

\newcommand{\usssPl}{\sss{\uPl}}

\newcommand{\Mp}{M_\usssPl}

\newcommand{\efolds}{$e$-folds}

\newcommand{\beq}{\begin{equation}}
\newcommand{\eeq}{\end{equation}}
\newcommand{\bea}{\begin{equation}\begin{aligned}}
\newcommand{\eea}{\end{aligned}\end{equation}}

\newlength{\wsingfig}
\newlength{\wdblefig}
\newlength{\wquadfig}
\newlength{\wtriplefig}
\newcommand{\Eq}[1]{Eq.~(\ref{#1})}
\newcommand{\Eqs}[1]{Eqs.~(\ref{#1})}
\newcommand{\Fig}[1]{Fig.~{\ref{#1}}}
\newcommand{\Figs}[1]{Figs.~{\ref{#1}}}
\newcommand{\Refa}[1]{Ref.~{\cite{#1}}}

\newcommand{\Sec}[1]{Sec.~\ref{#1}}

\newcommand{\App}[1]{Appendix~\ref{#1}}

\subheader{}

\title{Primordial black holes from a curvaton: the role of bimodal distributions}

\author[a,b]{Tomotaka Kuroda,}
\author[c,d,e]{Atsushi Naruko,}
\author[f]{Vincent Vennin,}
\author[b,a]{Masahide Yamaguchi}

\affiliation[a]{Department of Physics, Institute of Science Tokyo Tokyo, 152-8551, Japan}
\affiliation[b]{Cosmology, Gravity and Astroparticle Physics Group, Center for Theoretical Physics of the Universe, Institute for Basic Sclence (IBS), Daejeon, 34126, Korea}
\affiliation[c]{Center for Gravitational Physics and Quantum Information,
Yukawa Institute for Theoretical Physics, Kyoto University, Kyoto 606-8502, Japan}
\affiliation[d]{Asia Pacific Center for Theoretical Physics, Pohang 37673, Korea}
\affiliation[e]{School of Natural Sciences, National Institute of Technology (KOSEN), Gunma College, 580 Toriba, Maebashi, Gunma 371-8530, Japan}
\affiliation[f]{Laboratoire de Physique de l'Ecole Normale Sup\'erieure, ENS, CNRS, Universit\'e PSL, Sorbonne Universit\'e, Universit\'e Paris Cit\'e, 75005 Paris, France}

\emailAdd{kuroda.t.ad@m.titech.ac.jp} 
\emailAdd{naruko@yukawa.kyoto-u.ac.jp}
\emailAdd{vincent.vennin@phys.ens.fr}
\emailAdd{gucci@ibs.re.kr}

\date{today}

\begin{document}
\sloppy

\abstract{We investigate the formation of primordial black holes in curvaton models of inflation, where the curvature perturbation is not only generated by the inflaton but also by a light scalar field (the curvaton) that decays after inflation. During inflation, both fields are subject to quantum diffusion, owing to small-scale vacuum fluctuations crossing out the Hubble radius. After inflation, whether the curvaton dominates the universe or not depends on its field value when inflation ends. Since that value is stochastic, different regions of the universe undergo different post-inflationary histories. In practice, we show that this results in a double-peaked distribution for the number of \efolds~realised in these models. Since that number of \efolds~is related to the curvature perturbation by the $\delta N$ formalism, the presence of a second peak has important consequences for primordial black holes that we discuss. }

\arxivnumber{}

\begin{flushright}
YITP-25-49
\end{flushright}

\maketitle


\flushbottom


\section{Introduction}\label{Sec:1}
Cosmological inflation is a rapid expansion period in the very early universe. Quantum fluctuations generated during inflation are stretched by the accelerated expansion, successfully explaining observational phenomena such as the anisotropies of temperature fluctuations in the cosmic microwave background (CMB) \cite{Planck:2018vyg}. While these fluctuations are constrained to be small and quasi scale-invariant at large scales, as expected from a phase of single-field slow-roll inflation, at small scales they have not been measured yet and if their amplitude is large enough they could give rise to primordial black holes (PBHs). This would have important consequences. For instance, if their masses lie within the asteroid-mass range, PBHs are considered as a valid candidate for dark matter. Moreover, mergers of PBHs are expected to emit gravitational waves, which would leave a stochastic background of gravitational waves behind, possibly within the reach of future detectors.

PBHs form by gravitational collapse when fluctuations above a certain threshold re-enter the Hubble radius after inflation. Large-amplitude fluctuations cannot be properly described by conventional perturbation theory, hence non-perturbative approaches such as the stochastic-$\delta N$ formalism \cite{Starobinsky:1986fx, Fujita:2013cna, Vennin:2015hra} are often employed to describe non-linear super-Hubble dynamics during inflation. This formalism consists of a classical effective field theory for super-Hubble modes, which are subject to stochastic diffusion coming from quantum fluctuations in the sub-Hubble modes crossing out the Hubble radius. Combined with the $\delta N$ formalism~\cite{Starobinsky:1982ee, Starobinsky:1986fxa, Sasaki1996, Sasaki:1998ug, Lyth:2004gb}, this allows one to describe curvature perturbations as differences in the local amounts of inflation, as measured by the number of inflationary \efolds~$N=\ln(a)$, $a$ being the scale factor. In practice, the statistics of that number of \efolds~can be obtained by solving a first-passage-time problem~\cite{Vennin:2015hra,Pattison:2017mbe,Ezquiaga:2019ftu}, and is found to feature heavy non-Gaussian tails. This leads to enhanced probabilities to exceed the PBH formation threshold, which can be assessed using various techniques~\cite{Ando:2020fjm, Figueroa:2020jkf, Tada:2021zzj, Kitajima:2021fpq, Hooshangi:2021ubn, Gow:2022jfb, Raatikainen:2023bzk, Animali:2024jiz, Vennin:2024yzl, Jackson:2024aoo, Animali:2025pyf, Choudhury:2025kxg}.

As mentioned above, fluctuations at large scales are constrained to be small and quasi scale invariant, hence deviations from single-field slow-roll inflation must take place in the late stages of inflation in order for fluctuations to grow large. This requires either slow roll to be violated -- a typical example being ultra slow roll~\cite{Dimopoulos:2017ged} during which the curvature perturbations grow as the cube of the scale factor; or multiple-field effects to be become relevant. In this work, we consider the second possibility, and investigate one of the simplest setups where inflation is still driven by a single scalar field, the inflaton $\phi$, but in the presence of a light test field $\sigma$ dubbed curvaton~\cite{Linde:1996gt,Enqvist:2001zp,Lyth:2001nq,Moroi:2001ct}. Both the inflaton and the curvaton are assumed to be slowly rolling during inflation, after which they decay into radiation fluids. Such light test fields are expected in most models that have been proposed to
embed inflation in high-energy constructions~\cite{Linde:1993cn, Kawasaki:2000yn, Lyth:2002my, Kachru:2003sx, Allahverdi:2006we, Kallosh:2004yh, Martin:2013tda, Baumann:2014nda, Vennin:2015vfa, Moultaka:2016frs, Weymann-Despres:2023wly, Kaiser:2013sna, DeCross:2015uza}. This is why curvaton models are one of the most straightforward generalisations to the simplest inflationary setups where PBHs might form~\cite{Yokoyama:1995ex, Kawasaki:2012wr, Bugaev:2013vba, Ando:2017veq, Ando:2018nge, Chen:2019zza, Inomata:2020xad, Liu:2021rgq, Su:2025mam}.

Both $\phi$ and $\sigma$ can contribute to curvature perturbations at late time~\cite{Bartolo:2002vf,Dimopoulos:2003az,Langlois:2004nn,Lazarides:2004we,Moroi:2005np,Ellis:2013iea,Enqvist:2013paa,Byrnes:2014xua,Hardwick:2016whe,Torrado:2017qtr,Kumar:2019ebj}. In the limit where the curvaton is entirely responsible for curvature perturbations, non-Gaussianities are expected at the level $f_{\mathrm{NL}}=-5/4$~\cite{Valiviita:2006mz, Sasaki:2006kq}, where $f_{\mathrm{NL}}$ is the local non-Gaussianity parameter measuring the amplitude of the bispectrum in units of the power spectrum. The tail of the curvature distribution is however poorly described by the perturbative bispectrum, and how the presence of a curvaton affects to probability to realise large fluctuations, prone to forming PBHs, is the topic of ongoing investigations~\cite{Pi:2021dft, Stamou:2023vwz, Ferrante:2023bgz, Gow:2023zzp, Chen:2024pge, Stamou:2024xkk}.
 
In this work, we show how the statistics of curvature perturbations in curvaton models can be reconstructed with the stochastic-$\delta N$ formalism. We find that there exist regimes in parameter space where the distribution function of the curvature perturbation is bimodal. This is due to different reheating histories being selected by the curvaton field value, which fluctuates on the end-of-inflation hypersurface. This generic mechanism features a new type of non-Gaussian tails, which are not only characterised by a slower decay rate than Gaussian statistics, but by the appearance of a whole second maximum. This has important consequences for the formation of PBHs that we discuss.

The rest of the paper is organised as follows. \Sec{Sec:formalism} reviews the stochastic formalism in the presence of both {inflaton} and spectator fields and implements the $\delta N$ program to derive the distribution function of curvature perturbations on the curvaton decay surface. \Sec{Sec:4}  discusses the result and how it depends on the stochastic dynamics of the curvaton during inflation, with an emphasis on the consequences for PBH production. Our main results are presented in \Sec{Sec:5}, and the article ends with a few appendices to which some technical details are deferred. 

\section{The stochastic-$\delta N$ formalism in the presence of a curvaton}\label{Sec:formalism}

In this section, we briefly review the stochastic formalism, when spectator fields are explicitly included, and we implement the $\delta N$ program in the presence of a curvaton. Let us consider the case where inflation is driven by a set of inflaton fields collectively denoted as $\boldmathsymbol{\phi}$, in the presence of a set of curvaton fields collectively denoted as $\boldmathsymbol{\sigma}$. We assume that curvatons are spectator fields that are irrelevant to the inflationary dynamics, which is determined only by the inflatons. The action of the whole system comprises the Einstein-Hilbert action and the action of inflatons and curvatons,
\begin{align}
  S = \int \dd x^4 \sqrt{-g}\left(\frac{\Mp ^2}{2}R - \frac{1}{2}g^{\mu\nu} \partial_{\mu}\boldmathsymbol{\phi}\partial_{\nu}\boldmathsymbol{\phi} - V^{(\boldmathsymbol{\phi})} - \frac{1}{2}g^{\mu\nu}\partial_{\mu}\boldmathsymbol{\sigma}\partial_{\nu}\boldmathsymbol{\sigma} - V^{(\boldmathsymbol{\sigma})}    \right),
\end{align}
where $\Mp$ is the reduced Planck mass, $g^{\mu\nu}$ and $R$ are the (inverse) metric and its Ricci scalar, and $V^{(\boldmathsymbol{\phi})},\ V^{(\boldmathsymbol{\sigma})} $ are the potential of the inflatons and the curvatons respectively. They depend only on the inflaton and curvaton field values respectively, so the two sets of fields are not directly coupled. 

\subsection{Stochastic inflation with dynamical and spectator fields}
\label{sec:stochastic:inflation}

In the stochastic formalism, fields are split into an IR part, coarse-grained at a scale somewhat larger than the Hubble radius, and a UV part, which is the complement of the IR part. The UV part is described quantum mechanically using cosmological perturbation theory. When modes belonging to the UV part cross out the coarse-graining scale, they join the IR part whose classical dynamics is shifted by a stochastic noise that models this inflow of modes. In the separate-universe limit~\cite{Sasaki:1998ug,Wands:2000dp,Lyth:2003im,Lyth:2004gb}, each coarse-grained patch evolves independently, which implies that gradient interactions can be neglected in the IR dynamics.

In this limit, in the uniform-$N$ gauge, the IR field equations of motion read\footnote{Hereafter, for simpliciy, $\boldmathsymbol{\phi}$ and $\boldmathsymbol{\sigma}$ denote their IR parts only.}
 \begin{align}
  \frac{\dd\boldmathsymbol{\phi}}{\dd N} &= - \frac{V^{(\boldmathsymbol{\phi})}_{,\boldmathsymbol{\phi}}}{3H^2} + \frac{H}{2\pi}\xi_{\boldmathsymbol{\phi}},\label{Langevinslowroll1}\\
  \frac{\dd\boldmathsymbol{\sigma}}{\dd N} &= - \frac{V^{(\boldmathsymbol{\sigma})}_{,\boldmathsymbol{\sigma}}}{3H^2} + \frac{H}{2\pi}\xi_{\boldmathsymbol{\sigma}},\label{Langevinslowroll2}
\end{align}
where $\xi_{\boldmathsymbol{\phi}},\ \xi_{\boldmathsymbol{\sigma}}$ are independent normalised Gaussian noises. They are white if coarse-graining is performed through a step function in Fourier space, \ie 
\begin{align}
  \Braket{\xi_{X}(N)\xi_{X'}(N')}&= \delta_{X,X'}\delta(N-N'),
\end{align}
with $X$ and $X'$ denoting components of the fields $\boldmathsymbol{\phi}$ and $\boldmathsymbol{\sigma}$. The above equations are valid in the slow-roll regime, to which our analysis is restricted, although the stochastic formalism can be extended beyond the slow-roll attractor~\cite{Grain:2017dqa,Pattison:2019hef,Pinol:2020cdp,Cruces:2021iwq,Cruces:2024pni,Jackson:2024aoo}. In \Eqs{Langevinslowroll1}-\eqref{Langevinslowroll2}, $H=\dot{a}/a$ is the Hubble rate, it is related to the field values of the inflatons through the Friedmann equation
\begin{align}
\label{eq:Friedmann}
    H^2 = \frac{V^{(\boldmathsymbol{\phi})}}{3\Mp ^2}.
\end{align}
Moreover, inflation terminates when the first Hubble-flow parameter,
\begin{align}
  \epsilon_1 \equiv -\frac{\dd \ln H}{\dd N} = \frac{\Mp ^2}{2}\left(\frac{V^{(\boldmathsymbol{\phi})}_{,\boldmathsymbol{\phi}}}{V^{(\boldmathsymbol{\phi})}}\right)^2\, ,
\end{align}
crosses unity. 
From the Langevin equations~\eqref{Langevinslowroll1}-\eqref{Langevinslowroll2}, one can derive Fokker-Planck equations that drive the probability $P_{\mathrm{FP}}(\boldmathsymbol{\phi}, \boldmathsymbol{\sigma} ; N \vert \boldmathsymbol{\phi}_0, \boldmathsymbol{\sigma}_0 ; N_0)$ to find the fields at values $\boldmathsymbol{\phi}$ and $\boldmathsymbol{\sigma}$ at time $N$, given that they started from $\boldmathsymbol{\phi}_0$ and $\boldmathsymbol{\sigma}_0$ at time $N_0$. It is given by
\bea
\frac{\partial}{\partial N}P_{\mathrm{FP}}=
\Mp^2 \frac{\partial}{\partial\boldmathsymbol{\phi}} \left(\frac{V^{(\boldmathsymbol{\phi})}_{,\boldmathsymbol{\phi}}}{V^{(\boldmathsymbol{\phi})}} P_{\mathrm{FP}}  \right)
+\Mp^2 \frac{\partial}{\partial\boldmathsymbol{\sigma}} \left(\frac{V^{(\boldmathsymbol{\sigma})}_{,\boldmathsymbol{\sigma}}}{V^{(\boldmathsymbol{\phi})}} P_{\mathrm{FP}}  \right)
+\frac{1}{24\pi^2\Mp^2}\left(\frac{\partial^2}{\partial\boldmathsymbol{\phi}^2}+\frac{\partial^2}{\partial\boldmathsymbol{\sigma}^2}\right)\left( V^{(\bm{\phi})}P_{\mathrm{FP}}  \right)
\eea
where the arguments of $P_{\mathrm{FP}}$ are not written explicitly for conciseness.

An adjoint Fokker-Planck equation for the duration of inflation $\mathcal{N}_{\mathrm{inf}}$ can also be obtained. Since inflation is only driven by the inflatons, it only involves $\boldmathsymbol{\phi}$, and reads
\begin{align}
\frac{\partial }{\partial \mathcal{N}_{\mathrm{inf}}} \Pfpt (\mathcal{N}_{\mathrm{inf}};\boldmathsymbol{\phi}_*) =
\left(-\Mp^2\frac{V^{(\boldmathsymbol{\phi})}_{,\boldmathsymbol{\phi}}}{V^{(\boldmathsymbol{\phi})}} 
\frac{\partial}{\partial\boldmathsymbol{\phi}}+\frac{1}{24\pi^2\Mp^2} \frac{\partial^2}{\partial\boldmathsymbol{\phi}^2}\right)\Pfpt (\mathcal{N}_{\mathrm{inf}};\boldmathsymbol{\phi}_*). \label{adjointFokkerPlanckinflaton}
\end{align}
Here, $\Pfpt (\mathcal{N}_{\mathrm{inf}};\boldmathsymbol{\phi}_*)$ denotes the probability to realise $\mathcal{N}_{\mathrm{inf}}$ \efolds~of inflation starting from $\boldmathsymbol{\phi}_*$. In the $\delta N$ formalism~\cite{Starobinsky:1986fxa,Sasaki1996,Wands:2000dp,Lyth:2005fi}, fluctuations in $\mathcal{N}_{\mathrm{inf}}$ are nothing but the coarse-grained curvature perturbation on the end-of-inflation hypersurface, where the boundary condition $\Pfpt (\mathcal{N}_{\mathrm{inf}};\boldmathsymbol{\phi}_*)=\delta(\mathcal{N}_{\mathrm{inf}})$ needs to be imposed.

As will be made clear below, the curvature perturbation on the curvaton decay surface does not depend only on $\mathcal{N}_{\mathrm{inf}}$, but also on the curvaton values $\boldmathsymbol{\sigma}_\uend$ on the end-of-inflation hypersurface. The dynamics of the curvatons during inflation is rather involved, since the amplitude of the noise acting on them, $H/(2\pi)$, depends on the inflatons, which are themselves stochastic. There are however two limits where the gravitational interaction between the inflatons and the curvatons is negligible and the problem becomes tractable: if the inflaton fluctuations remain small and the noise amplitude can be computed along a reference trajectory, and if inflation proceeds at quasi constant $H$. In both cases, $H$ can be taken as an external, deterministic function of time, the joint distribution factorises, $P_{\mathrm{FP}}(\boldmathsymbol{\phi}, \boldmathsymbol{\sigma} ; N \vert \boldmathsymbol{\phi}_0, \boldmathsymbol{\sigma}_0 ; N_0)=P_{\mathrm{FP}}(\boldmathsymbol{\phi} ; N \vert \boldmathsymbol{\phi}_0 ; N_0)P_{\mathrm{FP}}( \boldmathsymbol{\sigma} ; N \vert \boldmathsymbol{\phi}_0, \boldmathsymbol{\sigma}_0 ; N_0)$ and the curvatons distribution follows the Fokker-Planck equation
\bea
\label{spectatorFokker-Planckeq}
\frac{\partial}{\partial N}P_{\mathrm{FP}}( \boldmathsymbol{\sigma} ; N \vert \boldmathsymbol{\phi}_0, \boldmathsymbol{\sigma}_0 ; N_0)= &
\frac{1}{3H^2(N)} \frac{\partial}{\partial\boldmathsymbol{\sigma}} \left[V_{,\boldmathsymbol{\sigma}}^{(\boldmathsymbol{\sigma})}P_{\mathrm{FP}}( \boldmathsymbol{\sigma} ; N \vert \boldmathsymbol{\phi}_0, \boldmathsymbol{\sigma}_0 ; N_0)\right] 
\\ &
+ \frac{H^2(N)}{8\pi^2}\frac{\partial^2}{\partial\boldmathsymbol{\sigma}^2}P_{\mathrm{FP}}( \boldmathsymbol{\sigma} ; N \vert \boldmathsymbol{\phi}_0, \boldmathsymbol{\sigma}_0 ; N_0) .
\eea
Here, the function $H(N)$ implicitly depends on the initial conditions $\boldmathsymbol{\phi}_0$ and $N_0$, which is why they appear as arguments of the curvatons distribution. When $H$ is constant, \Eq{spectatorFokker-Planckeq} admits a stationary solution given by
\begin{align}
    P_{\mathrm{stat}}(\boldmathsymbol{\sigma})\propto \exp\left(-\frac{8\pi^2 V^{(\boldmathsymbol{\sigma})}}{3H^4}\right).\label{stationarysolution}
\end{align}
Note however that for this distribution to be reached, the time scale over which $H$ varies, $N_H\sim 1/\epsilon_1$, needs to be much larger than the time $N_{\boldmathsymbol{\sigma}}$ over which the curvatons distribution function relaxes, $N_H \gg N_{\boldmathsymbol{\sigma}}$. For instance, if the curvatons have a mass $m_\sigma$ and do not self interact, then $N_{\boldmathsymbol{\sigma}}\sim H^2/m_\sigma^2$ and this condition is most often not fulfilled~\cite{Hardwick:2017fjo,Briaud:2023pky,Tokeshi:2024kuv}. This is why, although commonly employed, \Eq{stationarysolution} is of little use when it comes to assessing the statistics of the curvaton fields on the end-of-inflation hypersurface.

\subsection{The post-inflationary phase}
\label{sec:post:inflation}

After inflation, UV fluctuations stop crossing out the coarse-graining scale and the noises turn off. The dynamics of both $\boldmathsymbol{\phi}$ and $\boldmathsymbol{\sigma}$ becomes deterministic, and only depend on their field values on the end-of-inflation hypersurface, $\boldmathsymbol{\phi}_\uend$ and $\boldmathsymbol{\sigma}_\uend$. In the sudden-decay approximation~\cite{Malik:2006pm,Sasaki:2006kq}, fields decay when $H$ drops below a certain value that can be interpreted as their decay rate and that are usually denoted $\Gamma$ (there is one parameter $\Gamma$ per field). When the last field decays, the fluctuations in the total number of \efolds~$\mathcal{N}$ are related to the curvature perturbation, by virtue of the $\delta N$ formalism~\cite{Starobinsky:1982ee, Starobinsky:1986fxa, Sasaki1996, Sasaki:1998ug, Lyth:2004gb}. Denoting $N_{\mathrm{post}}$ the number of \efolds~realised between the end of inflation and the last field decay, the goal is thus to calculate the statistics of
\bea
\mathcal{N} = \mathcal{N}_{\mathrm{inf}} + N_{\mathrm{post}}\, .
\eea

As mentioned above, $N_{\mathrm{post}}$ only depends on $\boldmathsymbol{\phi}_\uend$ and $\boldmathsymbol{\sigma}_\uend$. Denoting $P( \mathcal{N}_{\mathrm{inf}},\boldmathsymbol{\phi}_\uend,\boldmathsymbol{\sigma}_\uend \vert \boldmathsymbol{\phi}_*,\boldmathsymbol{\sigma}_*)$ by the joint distribution of $\mathcal{N}_{\mathrm{inf}}$, $\boldmathsymbol{\phi}_\uend$ and $\boldmathsymbol{\sigma}_\uend$ produced during inflation, and conditioned to the initial field values $\boldmathsymbol{\phi}_*$ and $\boldmathsymbol{\sigma}_*$, one thus has
\bea
\label{eq:P:full}
P\left(\mathcal{N} \vert \boldmathsymbol{\phi}_*,\boldmathsymbol{\sigma}_*\right) = 
\int \dd  \mathcal{N}_{\mathrm{inf}} \dd \boldmathsymbol{\phi}_\uend \dd \boldmathsymbol{\sigma}_\uend P\left( \mathcal{N}_{\mathrm{inf}},\boldmathsymbol{\phi}_\uend,\boldmathsymbol{\sigma}_\uend \vert \boldmathsymbol{\phi}_*,\boldmathsymbol{\sigma}_*\right)
\delta\left[\mathcal{N}-\mathcal{N}_{\mathrm{inf}}-N_{\mathrm{post}}\left(\boldmathsymbol{\phi}_\uend,\boldmathsymbol{\sigma}_\uend\right)\right] .
\eea
In practice, if the curvature perturbation $\zeta$ is coarse-grained at a given scale, the distribution of $\zeta=\mathcal{N}-\langle \mathcal{N}\rangle$ can be obtained from the above where $\boldmathsymbol{\phi}_*$ and $\boldmathsymbol{\sigma}_*$ correspond to the field values when that scale crosses out the Hubble radius during inflation.\footnote{Strictly speaking, $\boldmathsymbol{\phi}_*$ and $\boldmathsymbol{\sigma}_*$ are stochastic variables, hence \Eq{eq:P:full} needs to be convolved with volume-weighted backward distribution functions~\cite{Ando:2020fjm, Animali:2024jiz, Animali:2025pyf}. Here we do not perform this convolution in order to better focus on the effects responsible for bimodal distributions and postpone its inclusion to future work.\label{footnote:backward:weighting}}

The joint distribution $P( \mathcal{N}_{\mathrm{inf}},\boldmathsymbol{\phi}_\uend,\boldmathsymbol{\sigma}_\uend \vert \boldmathsymbol{\phi}_*,\boldmathsymbol{\sigma}_*)$ depends on the model of inflation one considers. Hereafter, we assume that inflation is driven by a single inflaton $\phi$, hence in the slow-roll limit it ends at a unique field value $\bar{\phi}_\uend$, such that $\epsilon_1=1$. Under the approximation discussed above \Eq{spectatorFokker-Planckeq}, we also assume that the stochastic dynamics of the curvaton decouples from the one of the inflaton, hence
\bea
\label{eq:jointP:factorised}
P( \mathcal{N}_{\mathrm{inf}},{\phi}_\uend,{\sigma}_\uend \vert {\phi}_*,{\sigma}_*) = \Pfpt (\mathcal{N}_{\mathrm{inf}};{\phi}_*) \delta(\phi_\uend-\bar{\phi}_\uend)P_{\mathrm{FP}}({\sigma}_\uend ;  \mathcal{N}_{\mathrm{inf}} \vert {\phi}_*, {\sigma}_* ; 0).
\eea
Here, for simplicity, we have also assumed that there is a unique curvaton field $\sigma$. Inserting this factorised form into \Eq{eq:P:full}, one obtains
\bea
\label{eq:P:spec}
P(\mathcal{N}\vert\phi_*,\sigma_*) = \int \dd \sigma_\uend 
\Pfpt \left[\mathcal{N}-N_{\mathrm{post}}(\sigma_\uend);{\phi}_*\right]P_{\mathrm{FP}}\left[{\sigma}_\uend ; \mathcal{N}-N_{\mathrm{post}}(\sigma_\uend) \vert {\phi}_*, {\sigma}_* ; 0\right] .
\eea
One is left with a single integral over $\sigma_\uend$, where $\Pfpt $ can be obtained by solving \Eq{adjointFokkerPlanckinflaton}, $P_{\mathrm{FP}}$ by solving \Eq{spectatorFokker-Planckeq}, and the function $N_{\mathrm{post}}$ depends on the curvaton scenario. 

\subsection{A simple curvaton scenario}
\label{Sec:CurvatonScenarios}

\begin{figure}[t]
\centering 
\includegraphics[width=0.98\textwidth]{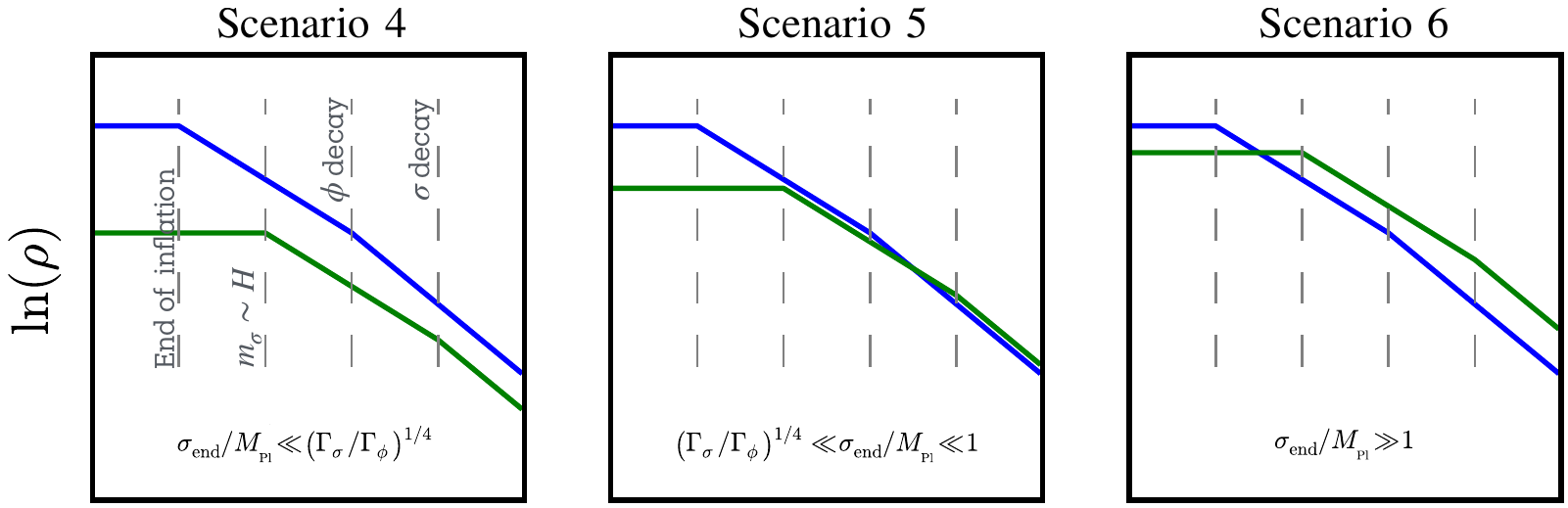} 
\caption{Curvaton scenarios considered in this work, for which $\Gamma_{\sigma}<\Gamma_{\phi}<m_{\sigma}<H_\uend$, according to the terminology of \Refa{Vennin:2015vfa}. The field value of the curvaton at the end of inflation, $\sigma_\uend$, determines whether and when the curvaton dominates the universe subsequently. The blue curves stand for the energy density stored in $\phi$ while the green curves are for $\sigma$.}
\label{fig:sketch} 
\end{figure}

As is common in curvaton cosmology, we assume that the curvaton is a free field with potential function $V^{(\sigma)}=m_\sigma^2\sigma^2/2$, where $m_\sigma\ll H_\uend$ for the curvaton to remain light during inflation. For explicitness, we consider the case where the decay rates of the inflaton and curvaton field, the mass of the curvaton and the inflationary Hubble rate are ordered according to
\bea
\Gamma_{\sigma}<\Gamma_{\phi}<m_{\sigma}<H_\uend\, .
\eea
Therefore, as $H$ decays after inflation ends, the curvaton first becomes massive and starts oscillating at the bottom of its potential, then the inflaton decays, and then the curvaton decays. We also assume that after inflaton ends, the inflaton oscillates at the bottom of the quadratic minimum of its potential, hence its energy density decreases as the one of matter until it decays, and subsequently as radiation. Likewise, the energy density contained in the curvaton decreases as matter after it becomes massive, and as radiation after it decays. 
In the generic classification of \Refa{Vennin:2015vfa} (see \Fig{fig:sketch}), this leaves us with three reheating scenarios, labelled as scenarios 4, 5 and 6, depending on the value of $\sigma_\uend$:
\begin{itemize}
\item scenario 4: $\sigma_\uend<\sigmaL\equiv(\Gamma_{\sigma}/\Gamma_{\phi})^{1/4}\Mp/K $ where $K=1/(\sqrt{2}e^{2/9})\simeq 0.566$. The curvaton never dominates the energy budget of the universe. 
\item scenario 5: $\sigmaL<\sigma_\uend<\sigmaR\equiv\Mp/K$. The curvaton starts dominating the energy budget of the universe between the inflaton decay and the curvaton decay.
\item scenario 6: $\sigma_\uend>\sigmaR$. The curvaton starts dominating the energy budget of the universe between the end of inflaton and the time it becomes massive.
\end{itemize}
Scenario 6 features a second phase of inflation driven by the curvaton~\cite{Dimopoulos:2011gb}, and thus slightly differs from the original curvaton proposal. In practice, it requires super-Planckian values for $\sigma_\uend$, and below we will work in regimes where such values have very low probabilities to be produced, which is why it will be enough to focus on scenarios 4 and 5.

In scenario 4, between the end of inflation and the decay of the inflaton, the universe behaves as matter dominated, $H^2\propto a^{-3}$, hence the number of \efolds~realised in that phase reads  
\begin{align}
    N_{\mathrm{end}\rightarrow \Gamma_{\phi}}^{(4)} = \frac{2}{3}\ln\left(\frac{H_{\mathrm{end}}}{\Gamma_{\phi}}\right) .
    \end{align}
Then, between the decay of the inflaton and the decay of the curvaton, the universe behaves as radiation dominated, $H^2\propto a^{-4}$, which leads to
\bea
N_{\Gamma_{\phi}\rightarrow \Gamma_{\sigma}}^{(4)} &= \frac{1}{2}\ln\left(\frac{\Gamma_{\phi}}{\Gamma_{\sigma}}\right) .
\eea
The total number of \efolds~is thus given by
\bea
N_{\mathrm{post}}^{(4)}=
N_{\mathrm{end}\rightarrow \Gamma_{\phi}}^{(4)}+N_{\Gamma_{\phi}\rightarrow \Gamma_{\sigma}}^{(4)}=
\frac{2}{3}\ln \left[\frac{H_{\mathrm{end}}}{\Gamma_{\sigma}}\left(\frac{\Gamma_{\sigma}}{\Gamma_{\phi}}\right)^{\frac{1}{4}}\right] .
\eea
This value does not depend on $\sigma_\uend$, since the curvaton never dominates the energy budget. 

In scenario 5, until the decay of the inflaton at $H=\Gamma_\phi$, the universe is dominated by the inflaton and its energy density behaves as that of pressureless matter. During this phase the curvaton remains frozen, $\sigma=\sigma_\uend$, until $H=m_\sigma$, and thereafter its energy density decreases like pressureless matter. After the inflaton decays, the universe is filled both with radiation from decay products of the inflaton and with pressureless matter, that is, the curvaton oscillating around the minimum of the potential. Then, as the universe expands, the matter component becomes dominant and eventually the curvaton decays at $H=\Gamma_\sigma$. The \efolds~realised in each phase are given by
\begin{align}
  N_{\mathrm{end}\rightarrow \Gamma_{\phi}} ^{(5)}= \frac{2}{3}\ln\left(\frac{H_{\mathrm{end}}}{\Gamma_{\phi}}\right),\quad N_{\Gamma_{\phi}\rightarrow \mathrm{eq}}^{(5)} = \frac{1}{2}\ln\left(\frac{\Gamma_{\phi}}{H_{\mathrm{eq}}}\right),\quad N_{\mathrm{eq}\rightarrow \Gamma_{\sigma}}^{(5)} = \frac{2}{3}\ln\left(\frac{H_{\mathrm{eq}}}{\Gamma_{\sigma}}\right),\label{efoldseqtogs}
\end{align}
where $H_{\mathrm{eq}}\sim K^4 \Gamma_{\phi}\frac{\sigma_{\mathrm{end}}^4}{\Mp^4}$ is the Hubble parameter {at the matter-radiation equality} between the inflaton decay and the curvaton decay. Adding them all together, we obtain
\begin{align}
  N_{\mathrm{post}}^{(5)} (\sigma_{\mathrm{end}}) = \frac{2}{3} \ln\left(\frac{H_{\mathrm{end}}}{\Gamma_{\sigma}}\frac{\sigma_{\mathrm{end}}}{\Mp }K\right). \label{Npost5}
\end{align}
One can check that, when $\sigma_\uend = \sigmaL$, $N_{\mathrm{post}}^{(4)}=N_{\mathrm{post}}^{(5)}$, as expected.

The two scenarios 4 and 5 thus allow us to split the integral~\eqref{eq:P:spec} according to
\bea
\label{eq:P:spec:cases}
& P(\mathcal{N}\vert\phi_*,\sigma_*) = \Pfpt \left[\mathcal{N}-N^{(4)}_{\mathrm{post}};{\phi}_*\right] \int_{\vert\sigma_\uend\vert<\sigmaL} \dd \sigma_\uend 
P_{\mathrm{FP}}\left[{\sigma}_\uend ; \mathcal{N}-N^{(4)}_{\mathrm{post}} \vert {\phi}_*, {\sigma}_* ; 0\right]
\\ &\quad\quad\quad
+ \int_{\sigmaL<\vert\sigma_\uend\vert<\sigmaR} \dd \sigma_\uend 
\Pfpt \left[\mathcal{N}-N^{(5)}_{\mathrm{post}}(\sigma_\uend);{\phi}_*\right]P_{\mathrm{FP}}\left[{\sigma}_\uend ; \mathcal{N}-N^{(5)}_{\mathrm{post}}(\sigma_\uend) \vert {\phi}_*, {\sigma}_* ; 0\right] ,
\eea
where we have used that, in scenario 4, $N_{\mathrm{post}}$ does not depend on $\sigma_\uend$.

Finally, with a quadratic potential the solution to the Fokker-Planck equation~\eqref{spectatorFokker-Planckeq}, with initial condition $P_{\mathrm{FP}}({\sigma} ; 0 \vert {\phi}_*, {\sigma}_* ; 0)=\delta(\sigma-\sigma_*)$, is given by the Gaussian distribution~\cite{Hardwick:2017fjo,Enqvist:2012xn}
\bea
\label{eq:FP:sol:Gaussian}
P_{\mathrm{FP}}\left({\sigma}; N \vert {\phi}_*, {\sigma}_* ; 0\right) = 
\frac{1}{\sqrt{2\pi \left\langle\delta \sigma^2(N)\right\rangle}}e^{-\frac{\left[\sigma-\left\langle\sigma(N)\right\rangle\right]^2}{2\left\langle\delta\sigma^2(N)\right\rangle}}
\eea
where
\begin{align}
    \Braket{\sigma(N)} = & \sigma_* \exp\left[-\frac{m_{\sigma}^2}{3}\int_{{0}}^{N}\frac{\dd N'}{H^{2}(N')}\right], \label{averagespec}\\
    \Braket{\delta\sigma^2(N)} = &  \int_{{0}}^{N}\dd N'\frac{H^{2}(N')}{4\pi^2}\exp\left[\frac{2m_{\sigma}^2}{3}\int_{N}^{N'}\frac{\dd N''}{H^{2}(N'')}\right] . \label{variancespec}
\end{align}
At late time, \ie when $N\gg N_{{\sigma}}\sim H^2/m_\sigma^2$, one recovers the stationary solution~\eqref{stationarysolution} if $H$ is constant. In this limit, $\sigma_\uend$ and $\mathcal{N}_{\mathrm{inf}}$ become independent variables in \Eq{eq:jointP:factorised}, hence the inflaton and curvaton fluctuations fully decouple.
In practice however, $N_{{\sigma}}$ is often much larger than the few tens of \efolds~that proceed between Hubble crossing of the scales of astrophysical interest and the end of inflation, hence the full solution~\eqref{eq:FP:sol:Gaussian}-\eqref{variancespec} must be kept.
\section{Bimodal distributions and primordial black holes}\label{Sec:4}
We are now in a position to combine previous results and compute the distribution function of $\mathcal{N}$ explicitly. The only missing ingredient is the first-passage-time distribution $\Pfpt({\mathcal{N}_\mathrm{inf}};\phi_*)$, which satisfies \Eq{adjointFokkerPlanckinflaton} whose solution depends on the single-field model of inflation one considers. As mentioned above, that distribution always possesses heavy, exponential tails, that are particularly relevant for the formation of PBHs. They are very few examples where the full first-passage-time distribution is known, but in the classical limit where $V^{(\phi)}\ll \Mp^4$ and $V^{(\phi)}\ll \Mp^4 \epsilon_1/\epsilon_2$ with $\epsilon_2\equiv \dd\ln(\epsilon_1)/(\dd N)$, it can be expanded~\cite{Vennin:2015hra, Pattison:2017mbe} and at leading order one recovers the Gaussian distribution predicted by cosmological perturbation theory,
\bea
\label{eq:Pfpt:Gaussian}
\Pfpt\left({\mathcal{N}_\mathrm{inf}};\phi_*\right) = \frac{1}{\sqrt{2\pi \left\langle\delta{\mathcal{N}_\mathrm{inf}}^2\right\rangle}}\exp\left[-\frac{\left({\mathcal{N}_\mathrm{inf}}-\left\langle{\mathcal{N}_\mathrm{inf}}\right\rangle\right)^2}{2 \left\langle\delta{\mathcal{N}_\mathrm{inf}}^2\right\rangle}\right]
\eea
where
\bea
\left\langle{\mathcal{N}_\mathrm{inf}}\right\rangle =  \int_{\phi_\uend}^{\phi_*} \frac{\dd\phi}{\Mp^2} \frac{V^{(\phi)}(\phi)}{V^{(\phi)}_{,\phi}(\phi)}
\quad\text{and}\quad
\left\langle\delta{\mathcal{N}_\mathrm{inf}}^2\right\rangle =\frac{1}{12\pi^2} \int_{\phi_\uend}^{\phi_*} \frac{\dd\phi}{\Mp^8} \frac{\left[V^{(\phi)}(\phi)\right]^4}{\left[V^{(\phi)}_{,\phi}(\phi)\right]^3}\, .
\eea
At this order, the non-Gaussian tails are not captured, but in what follows we will nonetheless use these expressions in order to focus on the non-Gaussianities that arise from the curvaton only. We therefore work in the regime where $V^{(\phi)}/\Mp^4$ is sufficiently small such that the non-Gaussian corrections to \Eq{eq:Pfpt:Gaussian} only affect parts of the distribution function of $\mathcal{N}$ that lie above the PBH formation threshold.
\subsection{Bimodal distributions}

\begin{figure}[t]
 \begin{center} 
  \subfigure[{${\phi_*}=5\Mp$}]{
   \includegraphics[width=.5\columnwidth]{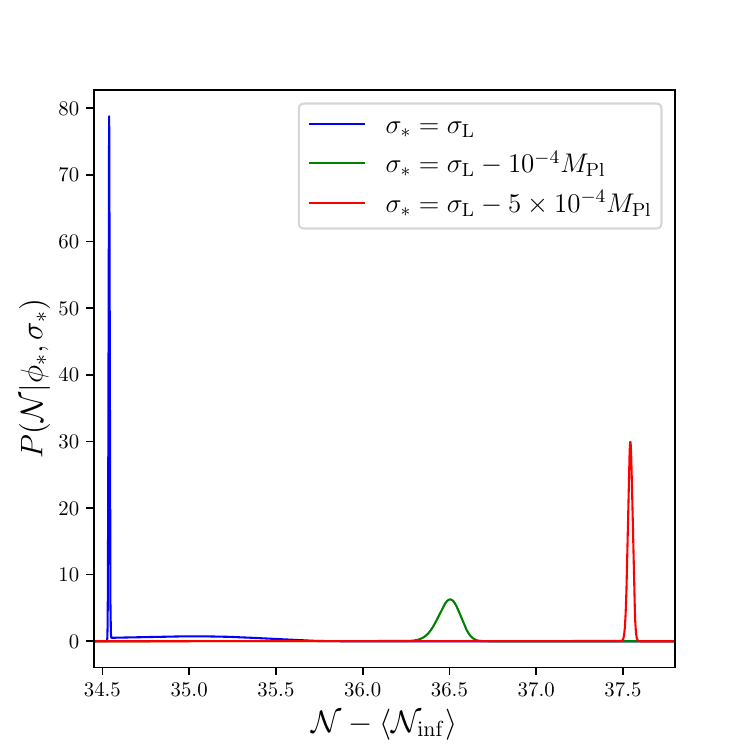}
  }~
  \subfigure[{${\sigma_*}=\sigmaL + 2\times10^{-5}\Mp$}]{
   \includegraphics[width=.5\columnwidth]{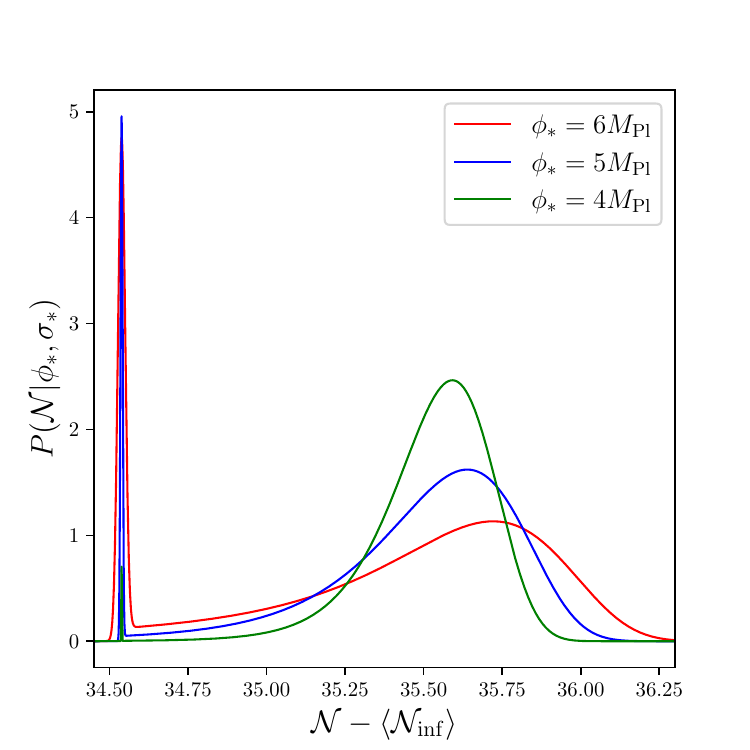}
  }
    \subfigure[{${\phi_*}=5\Mp$}]{
   \includegraphics[width=.5\columnwidth]{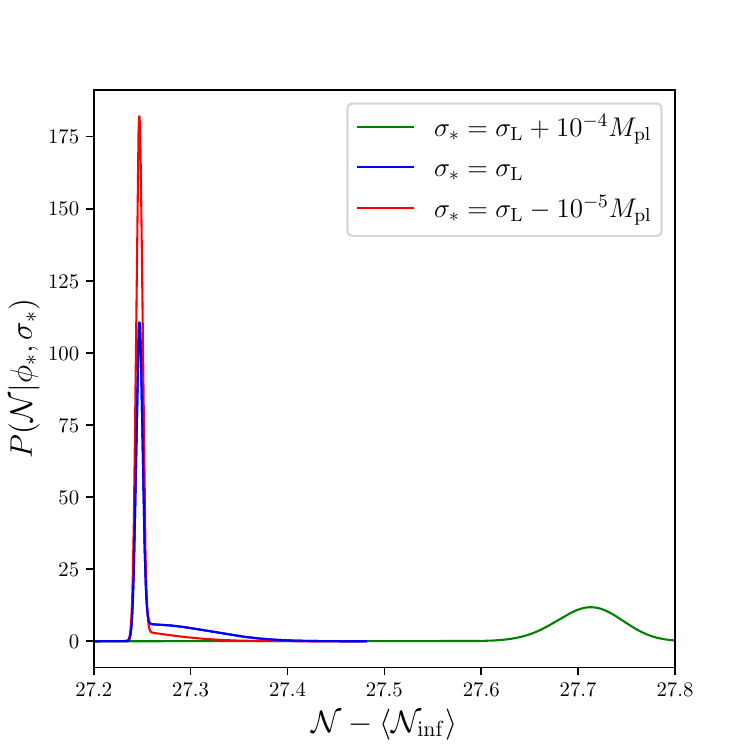}
  }~
  \subfigure[{${\sigma_*}=\sigmaL + 2.5\times10^{-5}\Mp$}]{
   \includegraphics[width=.5\columnwidth]{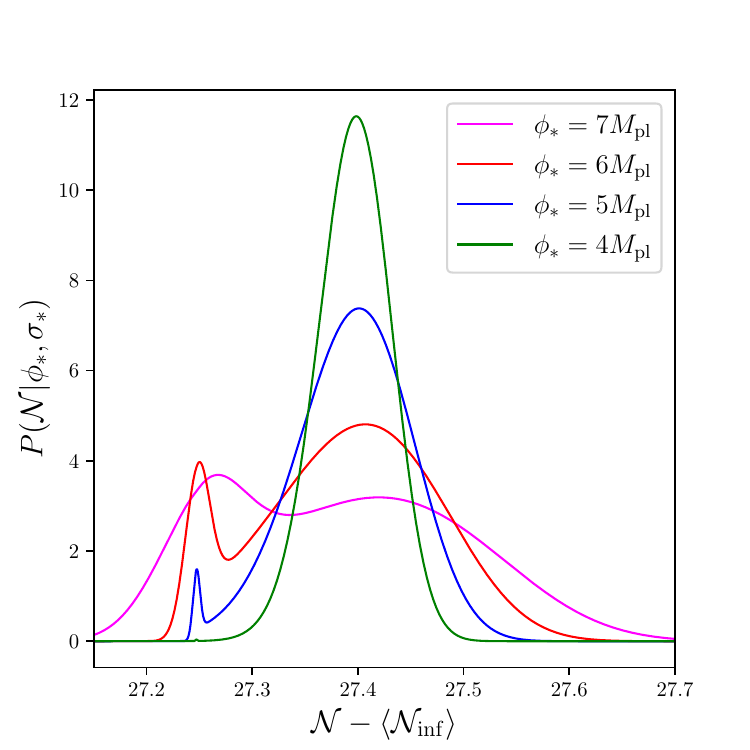}
   }
  \caption{Distribution functions for the total number of \efolds~$\mathcal{N}$ in the Starobinsky model of inflation in the presence of a curvaton with $H_\uend={10^{-5}}\Mp,\ m_\sigma =5 \times {10^{-9}}\Mp$; $\Gamma_\phi = {10^{-11}}\Mp$ and $\Gamma_\sigma = {10^{-33}}\Mp$ (upper panels); and  $\Gamma_\phi = {10^{-10}}\Mp$ and $\Gamma_\sigma = {10^{-27}}\Mp$ (lower panels); for various $\phi_*$ and $\sigma_*$. To ease comparison between different curves, which correspond to different scales in each panel, we show $\mathcal{N}-\langle\mathcal{N}_{\mathrm{inf}}\rangle$ on the horizontal axis, such that the parts of the distributions driven by the inflaton fluctuations are similarly centred. In the case represented in the right panels, at small scales (\ie low values of $\phi_*$), a secondary peak is produced, from which primordial black holes may form.} 
  \label{Fig:noneq}
 \end{center}
\end{figure}

In practice, we set the inflationary potential to the Starobinsky model~\cite{Starobinsky:1980te}
\bea
V^{(\phi)} = M^4\left(1-e^{-\sqrt{\frac{2}{3}}\frac{\phi}{\Mp}}\right)^2\, ,
\eea
and we let $H\simeq H_\uend$ in \Eq{averagespec}-\eqref{variancespec} since $H$ is almost constant in this plateau model. {Note that $M$ is a constant, related to the Hubble parameter $H$ via the Friedmann equation.} The distribution function of $\mathcal{N}$ is displayed in \Fig{Fig:noneq} for a few values of $\phi_*$ and $\sigma_*$, {using the explicit formulas given in \App{app:analytical:appr}.} {Recall that $\phi_*$ and $\sigma_*$ are scale dependent, hence different curves may be seen as describing different scales. However, for the sake of illustration, we have used a value of $H_\uend$ that is too large to accommodate the amplitude of temperature fluctuations in the CMB, otherwise the structure of the tail is difficult to display.}

The main observation is that for the parameters chosen in \Fig{Fig:noneq}, at small scales, \ie at low values of $\phi_*$, the distribution function acquires a second maximum at large $\mathcal{N}$. The two maxima can be traced back to the two contributions in the right-hand side of \Eq{eq:P:spec:cases}. The lower maximum comes from values of $\sigma_\uend<\sigmaL$ (scenario 4), for which $N_{\mathrm{post}}$ is a deterministic quantity and the fluctuations in $\mathcal{N}$ only come from those in $\mathcal{N}_{\mathrm{inf}}$, \ie from the inflaton fluctuations. This is why this maximum has a Gaussian shape, since we worked under the assumption of Gaussian first-passage-time distributions, see \Eq{eq:Pfpt:Gaussian}. The upper maximum comes from values of $\sigma_\uend>\sigmaL$, for which the fluctuations in $\mathcal{N}$ also receive a contribution from those in $N_{\mathrm{post}}(\sigma_\uend)$, \ie from the curvaton fluctuations. When varying $\phi_*$ and $\sigma_*$, \ie when changing the scale, one changes the relative size of these two contributions, hence the relative heights of the two maxima. We thus reach the important conclusion that there exist curvaton models where the curvature perturbation is dominated by a single Gaussian peak at large scales, in agreement with CMB observations, and possesses a secondary peak at small scales, from which PBHs may form.

\subsection{Sources to total curvature perturbations}\label{Sec:42}
\begin{figure}[htbp]
 \begin{center} 
  \subfigure{
   \includegraphics[width=.454\columnwidth]{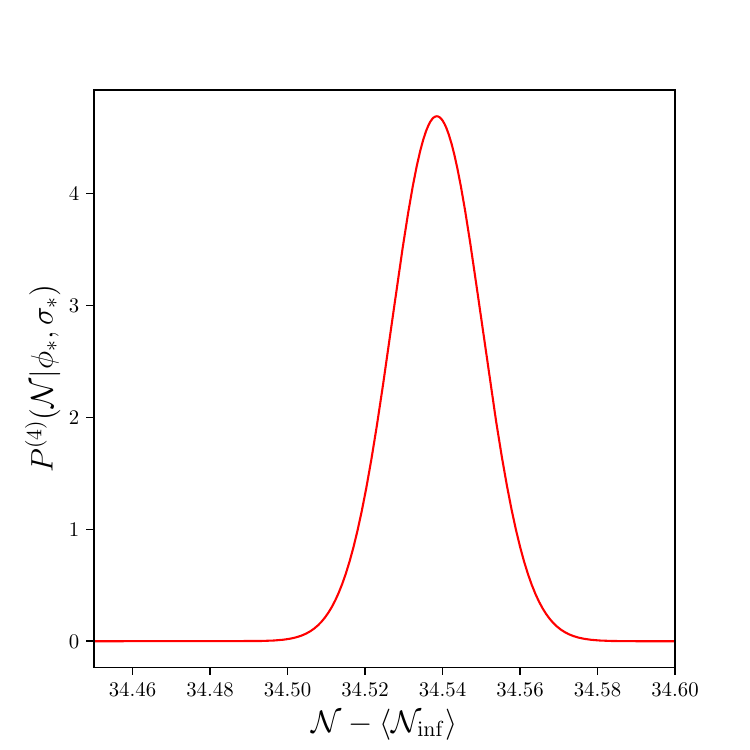}
  }~
  \subfigure{
   \includegraphics[width=.546\columnwidth]{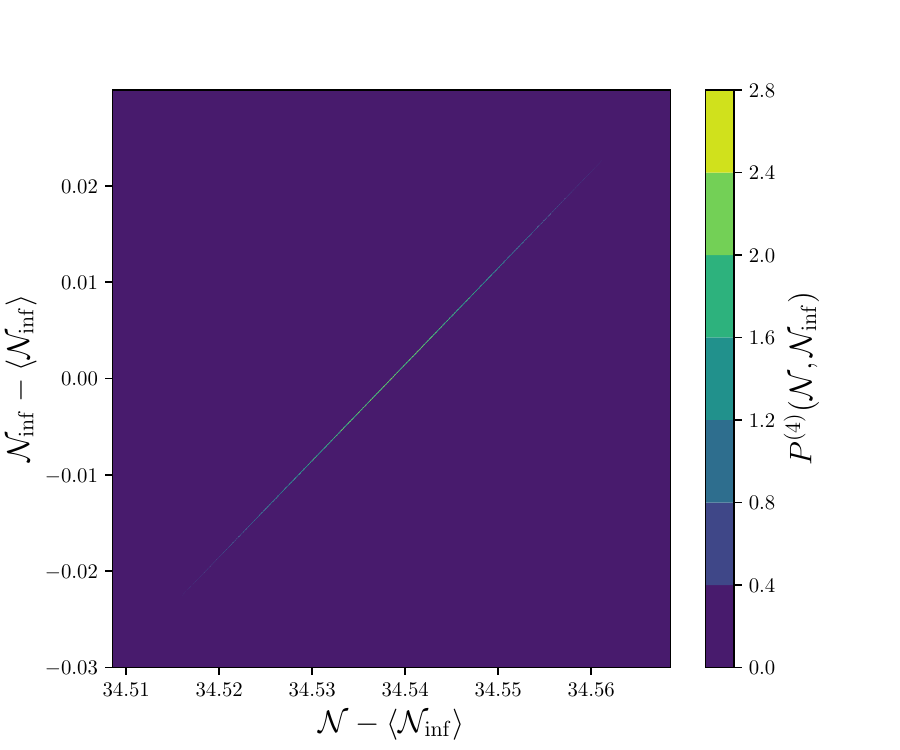}
  }\\
  \subfigure{
   \includegraphics[width=.454\columnwidth]{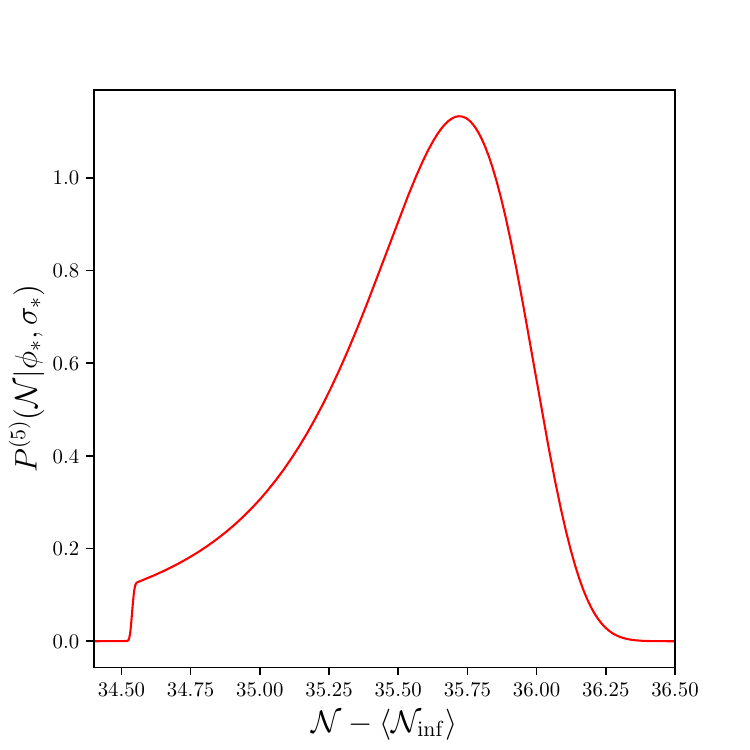}
  }~
  \subfigure{
   \includegraphics[width=.546\columnwidth]{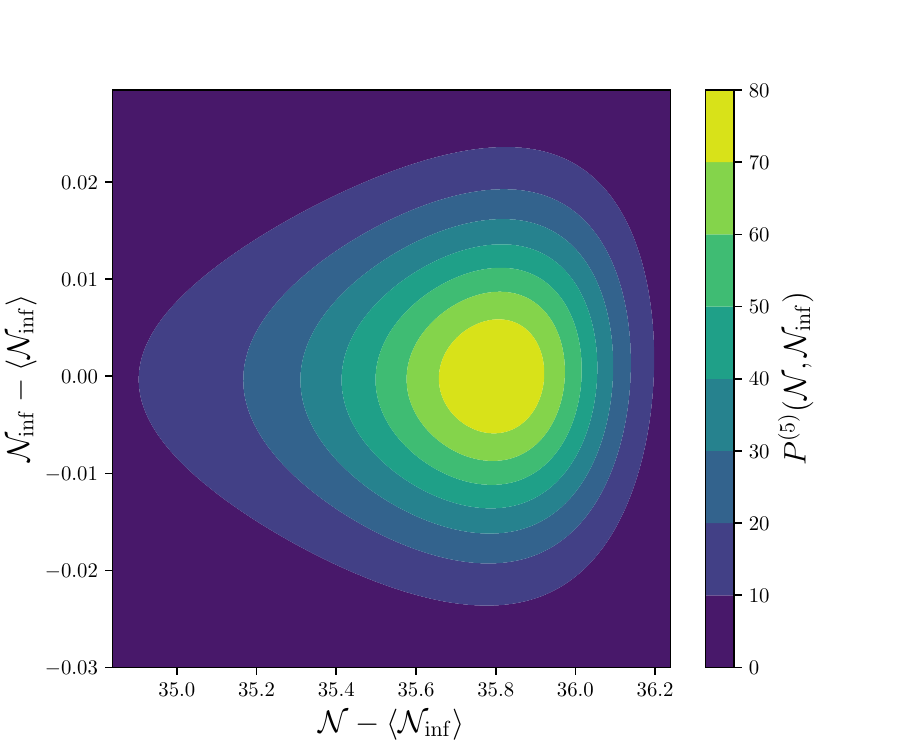}
  }
  \caption{Contributions from scenario 4 (upper panels) and scenario 5 (lower panels) to the distribution function of the total number of \efolds~$\mathcal{N}$ (left panels) and to the joint distribution function of $\mathcal{N}$ and $\mathcal{N}_{\mathrm{inf}}$ (right panels). The case being displayed corresponds to the red curve in the top right panel of \Fig{Fig:noneq} ($\sigma_*=\sigmaL + 2\times 10^{-5}\Mp$, $\phi_* = 6\Mp$, while other parameters are identical to \Fig{Fig:noneq}). 
  } 
  \label{Fig:eqjoint}
 \end{center}
\end{figure}

In order to confirm the interpretation proposed above that the two peaks in \Fig{Fig:noneq} correspond to the contributions from scenarios 4 and 5 respectively, hence from inflaton and curvaton fluctuations respectively, in this section we evaluate these two contributions separately. It is first worth stressing that $\mathcal{N}_{\mathrm{inf}}$ and $\sigma_\uend$ are not independent in general, since $\mathcal{N}_{\mathrm{inf}}$ appears as a parameter in the distribution function for $\sigma_\uend$. To make these correlations explicit, in the right panels of \Fig{Fig:eqjoint} we show the joint distribution 
\bea
P(\mathcal{N},\mathcal{N}_{\mathrm{inf}}) = & \int\dd\sigma_\uend
\Pfpt\left(\mathcal{N}_{\mathrm{inf}};\phi_*\right)
P_{\mathrm{FP}}\left[\sigma_\uend;\mathcal{N}-N_{\mathrm{post}}(\sigma_\uend)\vert\phi_*,\sigma_*;0\right]\\ &
\delta\left[\mathcal{N}-\mathcal{N}_{\mathrm{inf}}-N_{\mathrm{post}}\left( {\phi}_\uend, {\sigma}_\uend\right)\right] , \label{joint probability efold}
\eea 
which is nothing but the integrand in \Eq{eq:P:full}. The above can be decomposed into a contribution from scenarios 4 and 5, $P=P^{(4)}+P^{(5)}$, by splitting the integral over $\vert\sigma_\uend\vert<\sigmaL$ and  $\vert\sigma_\uend\vert>\sigmaL$ respectively, and this corresponds to the two rows in \Fig{Fig:eqjoint}.

One notices that 
$\mathcal{N}$ and $\mathcal{N}_{\mathrm{inf}}$ appear perfectly correlated in $P^{(4)}$. This is because, in scenario 4, $\sigma_\uend$ never dominates the energy of the universe and thus does not contribute to $\mathcal{N}$, whose fluctuations only receive contributions from those in $\mathcal{N}_{\mathrm{inf}}$. This confirms that the first peak arises from inflaton fluctuations only. In contrast, in $P^{(5)}$, the fluctuations in $\mathcal{N}$ are larger than those in $\mathcal{N}_{\mathrm{inf}}$ {(mind the scale of the axes)}, which means that the second peak receives contributions from curvaton fluctuations mainly. Notice however that the joint distribution is slightly more squeezed at low values of $\mathcal{N}$, which signals a small but finite amount of correlations between $\mathcal{N}$ and $\mathcal{N}_{\mathrm{inf}}$. This is because the width of the distribution of $\sigma_\uend$ depends on $\mathcal{N}_{\mathrm{inf}}$, hence the inflaton's and curvaton's fluctuations are partly correlated. In the left panels of \Fig{Fig:eqjoint}, we show the individual contributions from scenarios 4 and 5 to the distribution function of $\mathcal{N}$, where one can check that they correspond to the first and second peaks respectively as announced above.

In order to relate this discussion to standard results of curvaton cosmology, let us recall how the curvature perturbations is usually decomposed in this context~\cite{Lyth:2003im, Sasaki:2006kq}. When the curvaton decays, the inflaton behaves as radiation and the curvaton as pressureless matter, hence their curvature perturbations on hypersurfaces where they have uniform densities are given by
\bea
\zeta_\phi = \delta\mathcal{N} + \frac{1}{4}\ln\left(\frac{\rho_\phi}{\bar{\rho}_\phi}\right)
\quad\text{and}\quad
\zeta_\sigma = \delta\mathcal{N} + \frac{1}{3}\ln\left(\frac{\rho_\sigma}{\bar{\rho}_\sigma}\right)
\eea
respectively. Here, $\rho_\phi$ and $\rho_\sigma$ are defined on uniform total energy density hypersurfaces, and $\bar{\rho}_\phi$ and $\bar{\rho}_\sigma$ are their mean values. The total density is uniform on the curvaton decay hypersurface $H=\Gamma_\sigma$, $\rho_\phi+\rho_\sigma=\bar{\rho}$, which gives rise to
\bea
\bar{\rho}_\phi e^{4(\zeta_\phi-\delta\mathcal{N})}+\bar{\rho}_\sigma e^{3(\zeta_\sigma-\delta\mathcal{N})} = \bar{\rho}_\phi+\bar{\rho}_\sigma\, .
\eea
This allows one to non-linearly relate $\delta\mathcal{N}$ to $\zeta_\phi$ and $\zeta_\sigma$. For instance, at linear order, one finds
\bea
\label{eq:deltaN:zetaphi:zetasigma}
\delta\mathcal{N} = r_{\mathrm{dec}}\zeta_\sigma + \left(1-r_{\mathrm{dec}}\right) \zeta_\phi
\quad\text{where}\quad
r_{\mathrm{dec}} = \frac{3\bar{\rho}_\sigma}{3\bar{\rho}_\sigma+4\bar{\rho}_\phi}\, ,
\eea
which makes explicit that the curvature perturbation receives contributions from both inflaton and curvaton fluctuations. Let us also note that, with the approximations adopted in this work, the density fluctuations of the subdominant field at the time of curvaton decay are not accounted for. This explains why the curvaton fluctuations do not contribute in scenario 4. In practice, this is valid only when $\sigma_\uend\ll\sigmaL$, since $r_{\mathrm{dec}}\ll 1$ in that limit, but the approximation becomes inaccurate at $\sigma_\uend\lesssim\sigmaL$ where curvaton fluctuations are known to generate large non-Gaussianities even in scenario 4~\cite{Vennin:2015vfa}. These would increase the probability to realise large fluctuations, hence by non considering them we are working under conservative assumptions, although their inclusion constitutes an important direction for future work.

\subsection{Comments on primordial black holes}\label{Sec:422}

The presence of a secondary peak is clearly relevant for the formation of extreme objects such as PBHs~\cite{Nakama:2016kfq}. They constitute a particular type of ``heavy'' tails, which is not only characterised by a slower suppression rate than Gaussian statistics, but by the appearance of another maximum. This substantially increases the probability to realise large fluctuations, prone to forming PBHs. In \Refa{Shinohara:2021psq}, it is also shown that PBHs arising from secondary peaks are highly clustered, which has important consequences for their subsequent merger rate and for observational signatures in general.

\begin{figure}[t]
 \begin{center} 
  \subfigure[$H_\uend={10^{-5}}\Mp$, $m_\sigma =5 \times {10^{-9}}\Mp$,\newline $\Gamma_\phi = {10^{-11}}\Mp$, $\Gamma_\sigma = {10^{-33}}\Mp$.]{
   \includegraphics[width=.5\columnwidth]{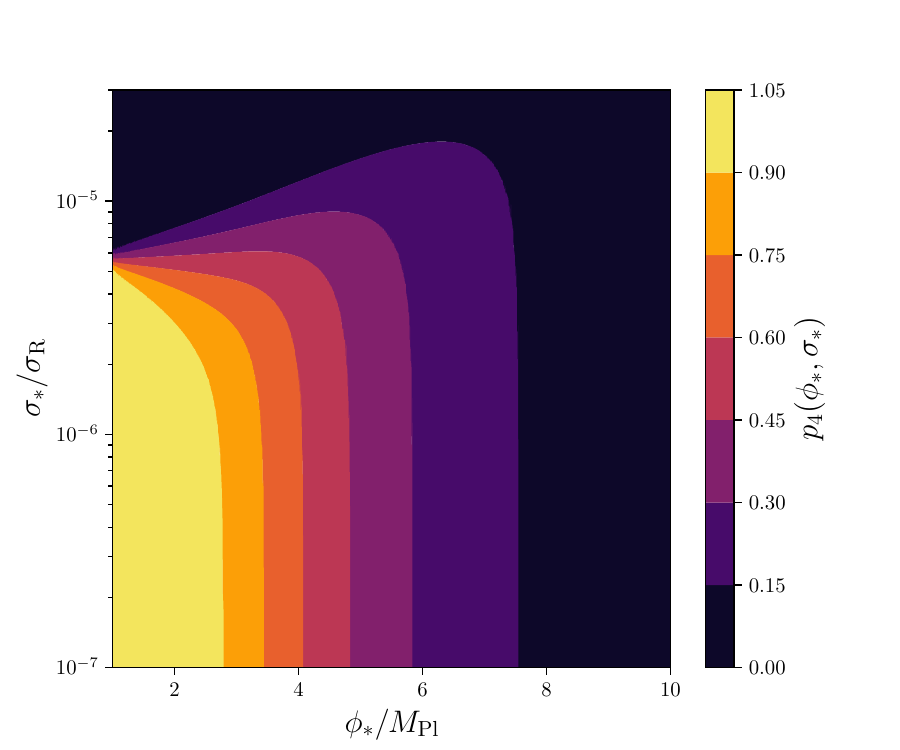}
  }~
  \subfigure[$H_\uend={10^{-6}}\Mp$, $m_\sigma = {10^{-9}}\Mp$,\newline $\Gamma_\phi = {10^{-13}}\Mp$, $\Gamma_\sigma = {10^{-17}}\Mp$.]{
   \includegraphics[width=.5\columnwidth]{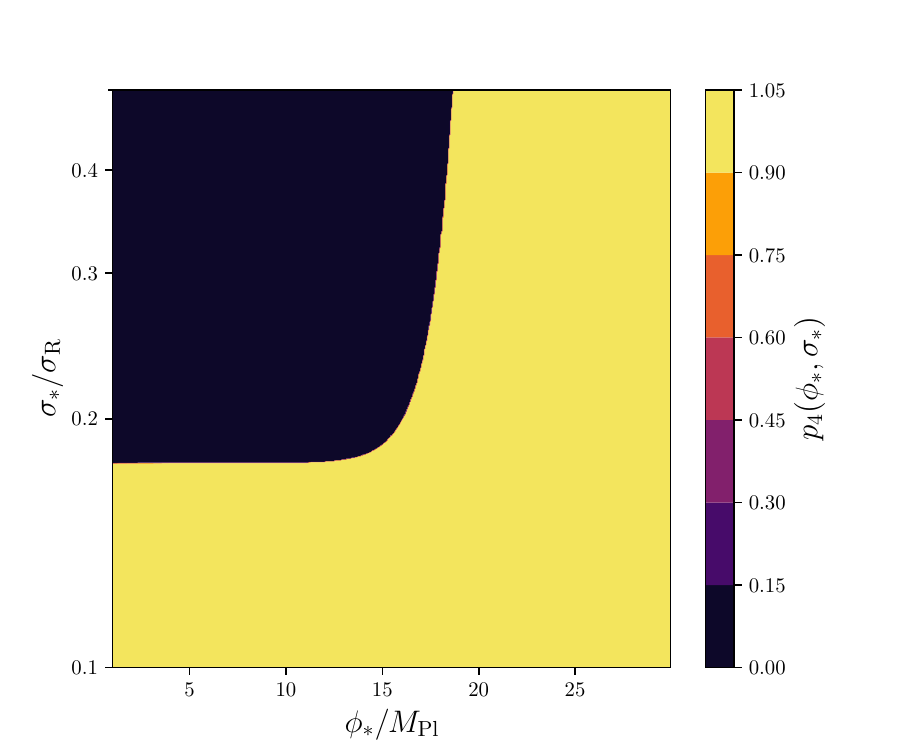}
  }
  \caption{{Probability $p_4$ of realising the post-inflationary scenario 4 (see \Fig{fig:sketch}). When $p_4\simeq 1$, the curvature perturbation is dominated by its first peak, while for $p_4$, the second peak is larger. The relative amplitudes of these peaks depend on $\phi_*$ and $\sigma_*$, hence on the scale at which the curvature perturbation is coarse-grained.}} 
  \label{Fig:p4}
 \end{center}
\end{figure}

The goal of the present work is not yet to characterise the abundance of PBHs precisely, but let us try and identify the region in parameter space where they form. As a rule of thumb, the relative weight between the first and the second peak is given by the relative probabilities to fall in scenarios 4 or 5. The probability $p_4$ associated to scenario 4 is the probability for $\vert\sigma_\uend\vert$ to be smaller than $\sigmaL$, hence from \Eq{eq:FP:sol:Gaussian}
\bea
p_4 = \frac{1}{2}\left[\mathrm{erf}\left(\frac{\sigmaL-\langle\sigma_\uend\rangle}{\sqrt{2 \langle\delta\sigma^2_\uend\rangle}}\right)+\mathrm{erf}\left(\frac{\sigmaL+\langle\sigma_\uend\rangle}{\sqrt{2 \langle\delta\sigma^2_\uend\rangle}}\right)\right]\, .
\eea
Approximating $H\simeq H_\uend$, and $\mathcal{N}_{\mathrm{inf}}\sim N_*(\phi_*)=\Braket{\mathcal{N}_\mathrm{inf}(\phi_*)}$, the classical number of inflationary \efolds~realised from $\phi_*$, one can replace $\langle\sigma_\uend\rangle = \sigma_*e^{-m_\sigma^2 N_*/(3 H_\uend^2)}$ and $\langle\delta\sigma^2_\uend\rangle = 3H_\uend^4/(8\pi^2m_\sigma^2)[1-e^{-2m_\sigma^2 N_*/(3H_\uend^2)}]$ in the above, and get an estimate for $p_4(\phi_*,\sigma_*)$. This function is displayed in \Fig{Fig:p4}.
{As inflation proceeds, $\phi_*$ decays, hence for $p_4$ to be smaller at larger scales (in order for the second peak to be larger), one requires $p_4$ to increase with $\phi_*$. This is the case in the right panel but not in the left panel. This implies that, to achieve a scenario where the second peak becomes important at small scales but leaves large scales unmodified, the right set of parameters needs to be chosen, and a more systematic exploration of parameter space is required.}

\section{Conclusion}\label{Sec:5}

In this work, we have investigated the formation of primordial black holes in curvaton models of inflation. Making use of the stochastic-$\delta N$ formalism, we have shown how the distribution function for the number of \efolds~$\mathcal{N}$ elapsed during inflation and until the last field decays can be computed. Since fluctuations in $\mathcal{N}$ are related to the curvature perturbations $\zeta$, this allows one to reconstruct the statistics of $\zeta$ in a non-perturbative way.

We have applied our formalism to a simple curvaton model where, after inflation, the curvaton becomes heavy, then the inflaton decays, and then the curvaton decays. Changing the ordering between these events would lead to other types of curvaton scenarios~\cite{Vennin:2015vfa} but similar considerations would then apply. During inflation, both the inflaton and the curvaton are subject to quantum diffusion, coming from sub-Hubble scales crossing out the Hubble radius and acting as a stochastic noise. After inflation, depending on the value of the curvaton at the end of inflation, the curvaton may or may not dominate the energy budget of the universe, which gives rise to different numbers of post-inflationary \efolds. 

Since the curvaton field value at the end if inflation is a random variable, whether the curvaton dominates after inflation or not is also stochastic. In other words, different regions of space reheat through different scenarios, hence undergo different numbers of post-inflationary \efolds. This leads to bimodal distributions for the total number of \efolds, which we have computed under several simplifying assumptions. We have found that these distributions generally contain a first quasi-Gaussian peak, arising mostly from inflaton fluctuations, and a secondary peak at larger $\mathcal{N}$, mostly coming from curvaton fluctuations. The relative size between these two peaks depends on the initial field values from which the stochastic dynamics is integrated during inflation, hence on the scales at which the curvature perturbation is coarse grained. This suggests the tantalising possibility~\cite{Shinohara:2021psq} that at large scales, curvature perturbations are dominated by the first Gaussian peak, in agreement with CMB observations, while the secondary peak becomes important at smaller scales and triggers the formation of large density fluctuations, likely to collapse into primordial black holes. 

Let us note that to reach this conclusion, a number of simplifying assumptions was performed, in order to focus on the physical effects we wanted to highlight. We do not expect that relaxing these approximations would alter our conclusions, but it might be necessary to reach robust quantitative estimates of the PBH abundance in curvaton models. In particular, we have neglected gravitational coupling between the inflaton and the curvaton fields during inflation, as well as the non-Gaussianities in the inflaton's contribution to the curvature perturbation (namely the exponential tails of the first-passage-time distribution). As discussed around \Eq{eq:deltaN:zetaphi:zetasigma}, we have also neglected the contribution from density fluctuations of the subdominant field across the final decay surface, which are expected to generate large non-Gaussianities~\cite{Vennin:2015vfa} close to the parameter-space boundary between different reheating scenarios. Moreover, we have worked in regimes where contributions from the inflating-curvaton channel (labelled as scenario 6 above) are negligible, although they may become relevant at larger curvaton field values. These effects should further enhance the relevance of the non-Gaussian tails in curvaton models and are therefore worth investigating further. 
Finally, we have derived the distribution function of $\mathcal{N}$ conditioned on $\phi_*$ and $\sigma_*$, the field values at Hubble crossing of the scale over which $\mathcal{N}$ is coarse grained. As mentioned in footnote~\ref{footnote:backward:weighting}, these field values are however stochastic, and follow backward distributions against which our result should be marginalised. Since the curvaton field value is not fixed on the end-of-inflation hypersurface (contrary to the inflaton's field value), these backward distributions depend on $\mathcal{N}$ itself, which should introduce an additional source of correlations between inflaton and curvaton fluctuations that we plan to study in the future.

\section*{Acknowledgments}
We thank Diego Cruces and Junsei Tokuda for useful discussions.
 T.K. acknowledges financial support from the Institute for Basic Science (IBS) under the project code, IBS-R018-D3. The work of A.N. was partly supported by JSPS KAKENHI Grant Numbers 20H05852, JP19H01891, JP23H01171, 23K25868. 
 A.N. thanks the molecule workshops ``Revisiting cosmological non-linearities in the era of precision surveys” YITP-T-23-03 and ``Extreme Mass Dark Matter Workshop: from Superlight to Superheavy” YITP-T-23-04 since discussions during the workshop were useful for this work.  
 M.Y. is supported by IBS under the project code, IBS-R018-D3, and by JSPS Grant-in-Aid for Scientific Research Number JP23K20843.

\appendix

\section{Steepest descent method}\label{Sec:A}
In this appendix, let us introduce the steepest-descent method, which will be employed to approximate the contribution from scenario 5,  $P^{(5)}(\mathcal{N};\phi_*,\sigma_*)$, in \App{app:analytical:appr}. The second term in \Eq{eq:P:spec:cases} is of the form
\begin{align}
    &P^{(5)}(\mathcal{N}\vert \phi_*,\sigma_*)\supset \frac{1}{\sqrt{2\pi\epsilon} } \exp\left[- \frac{\left(\mathcal{N} - \Braket{\mathcal{N}_{\mathrm{inf}}}\right)^2}{2\epsilon}\right] \int_{\sigmaL}^{\sigmaR} \dd\sigma_{\mathrm{end}} \exp\left[\frac{f(\sigma_\mathrm{end})}{\epsilon} \right] g(\sigma_\mathrm{end})  \label{NLOoriginal},
  \end{align}
  where 
  \begin{align}
    \epsilon &\equiv \Braket{\delta{\mathcal N_\mathrm{inf}}^2}\\
      f(\sigma_\mathrm{end}) &\equiv \left(\mathcal{N} - \Braket{\mathcal{N}_{\mathrm{inf}}}\right)  N_\mathrm{post}^{(5)}(\sigma_\mathrm{end}) - \frac{1}{2}\left[N_\mathrm{post}^{(5)}(\sigma_\mathrm{end})\right]^2, \\
      g(\sigma_\mathrm{end}) &\equiv P_{\mathrm{FP}}\left[{\sigma}_\mathrm{end} ; \mathcal{N}-N^{(5)}_{\mathrm{post}}(\sigma_\mathrm{end}) \vert {\phi}_*, {\sigma}_* ; 0\right] .\label{g}
  \end{align}
In the classical limit where inflationary fluctuations are small, $\epsilon\ll 1$, and since $f$ has one local maximum at $\sigma_{0} \equiv \sigmaL \exp[\frac{3}{2}(\mathcal{N} - N^{(4)}_{\mathrm{post}} - \Braket{\mathcal{N}_{\mathrm{inf}}})]$ (and diverges to minus infinity at the end point of the integration domain) the integrand is strongly peaked around $\sigma_0$. The result for the integral thus depends on whether $\sigma_0$ lies in the integration domain or not, and there are thus three cases to distinguish: $\sigmaL < \sigma_{0} < \sigmaR$, $\sigma_{0} < \sigmaL$, and $\sigma_{0} > \sigmaR$. Note that the situation is not trivial near the boundary, \ie when $\sigma_{0} \sim \sigma_{\mathrm{L}(\mathrm{R})}$ , which means that $f'(\sigma_{\mathrm{L}(\mathrm{R})})\sim 0$, where $\mathcal{N}$ does not satisfy 
\begin{align}
  \xi_{\sigma_\mathrm{end}}(\mathcal{N}) & \equiv \left|\frac{f''(\sigma_\mathrm{end})x^2}{2f'(\sigma_\mathrm{end})x}\right|_{x=\epsilon/[-f'(\sigma_\mathrm{end})]}
  = \left| \frac{3}{2} \epsilon \frac{\mathcal{N} - N_{\mathrm{post}}^{(4)} - \Braket{\mathcal{N}_{\mathrm{inf}}} 
  - \frac{2}{3}}{\left(\mathcal{N} -N_{\mathrm{post}}^{(4)} -\Braket{\mathcal{N}_{\mathrm{inf}}} \right)^2 }\right|\ll1,
\end{align}
which is defined as the ratio between the first and the second term in the Taylor expansion of $f$ around $\sigma_{\mathrm{end}}$([see \Eq{firstderapp1} below). In the following, $\sigma$ is used instead of $\sigma_{\mathrm{end}}$ to light up notations.

\subsection*{$(\mathrm{i})\ \sigmaL < \sigma_{0} < \sigmaR$}
The integration \eqref{NLOoriginal} is rewritten as 
\begin{align}
    \int_{\sigmaL}^{\sigmaR} \dd\sigma e^{f(\sigma)/\epsilon} g(\sigma) = \left(\int_{-\infty(0)}^{\infty} - \int_{\sigmaR}^{\infty} - \int^{\sigmaL}_{-\infty(0)}\right) \dd\sigma e^{f(\sigma)/\epsilon} g(\sigma) . \label{A.2}
\end{align}
By the steepest-descent method, the first integral can be expanded according to, 
\begin{align}
   &\int_{-\infty}^{\infty} \dd\sigma e^{f(\sigma)/\epsilon} g(\sigma) \notag\\
   &\quad = e^{f_0/\epsilon} \sqrt{\frac{2\pi \epsilon}{- f''_0}} \left\{ g_0 + \left[\left(\frac{g'_0f'''_0}{3!} + \frac{g_0f''''_0}{4!}\right)\frac{3}{f''^2_0} + \frac{g''_0}{2! (-f''_0)}\right]\epsilon + \mathcal{O}(\epsilon^{2}) \right\},\label{steepest}
\end{align}
where $f_0 \equiv f(\sigma_0),\ g_0 \equiv g(\sigma_0)$. This result is independent of $\xi_{\sigma}$, while the second and third integrals can be computed in accordance with $\xi_{\sigma}$ in the following manner. At first, let us focus on the case where $\mathcal{N}$ satisfies $\xi_\sigma \ll 1$. The second integral also can be calculated with relative ease, because only the first order of Taylor expansion should be taken into account due to $\xi_{\sigmaR} \ll 1$,
 \begin{align}
    \int_{\sigmaR}^{\infty} \dd\sigma e^{f(\sigma)/\epsilon} g(\sigma) 
    &= \int_{\sigmaR}^{\infty} \dd \sigma \exp\left[\frac{f_\mathrm{R}}{\epsilon} + \frac{f'_\mathrm{R}}{\epsilon} (\sigma-\sigmaR) + \frac{f''_\mathrm{R}}{2!\epsilon} (\sigma-\sigmaR)^2 + \cdots\right] [g_\mathrm{R} + g'_\mathrm{R} (\sigma-\sigmaR) + \cdots] \notag\\
    &= e^{f_\mathrm{R}/\epsilon}\int_0^{\infty} \dd x e^x \left[1+ \epsilon \frac{f_\mathrm{R}''}{2(f'_\mathrm{R})^2} x^2 +\cdots\right] 
    \left[ g_\mathrm{R} + \epsilon \frac{g_\mathrm{R}'}{f_\mathrm{R}'} x + \cdots \right] \notag\\
    &= e^{f_\mathrm{R}/\epsilon}  \left[ \frac{g_\mathrm{R}}{(-f_\mathrm{R}')}\epsilon  + \mathcal{O}(\epsilon^2)\right], \label{firstderapp1}
\end{align}
where $x\equiv (\sigma - \sigmaR) f_\mathrm{R}'/\epsilon $, $f_\mathrm{R} \equiv f(\sigmaR),\ g_\mathrm{R} \equiv g(\sigmaR) $. In this computation, the value of the integral is dominated by $e^{f(\sigma)/\epsilon}$, and a large contribution comes from $0<x< \epsilon/(-f_\mathrm{R}')$. Then, it is necessary that $\xi_{\sigmaR}\equiv\left|f''x^2/(2f'x)\right|_{x=\epsilon/(-f_\mathrm{R}')}\ll1$ in order for this computation to be justified, as mentioned above. Otherwise, higher-order terms must be taken into account. The third integral is also calculable in the same manner, 
 \begin{align}
    \int^{\sigmaL}_{-\infty} \dd\sigma e^{f(\sigma)/\epsilon} g(\sigma) 
    &= e^{f_\mathrm{L}/\epsilon}  \left[ \frac{g_\mathrm{L}}{f_\mathrm{L}'}\epsilon  + \mathcal{O}(\epsilon^2)\right],\label{firstderapp2}
\end{align}
where $f_\mathrm{L} \equiv f(\sigmaL),\ g_\mathrm{L} \equiv g(\sigmaL)$. Combining these, the final expression is obtained up to $e^{f/\epsilon} \mathcal{O}(\epsilon)$, 
\begin{align}
    &e^{f_0/\epsilon} \sqrt{\frac{2\pi \epsilon}{- f''_0}} g_0 -  e^{f_\mathrm{R}/\epsilon} \sqrt{\epsilon} \left( \frac{g_\mathrm{R}}{-f'_\mathrm{R}}\epsilon^{1/2}\right) -  e^{f_\mathrm{L}/\epsilon} \sqrt{\epsilon} \left( \frac{g_\mathrm{L}}{f'_\mathrm{L}}\epsilon^{1/2}\right),
\end{align}
which leads to \Eq{NLOneq4} below.

Next, let us focus on the $\xi_{\sigma}\not\ll 1$ case where the first derivative of the Taylor expansion of $f$ is relatively small, and the second derivative must be considered when calculating the second and third integrals in \Eq{A.2}. There are two possibilities, $\xi_{\sigmaR}\not\ll 1$ and $\xi_{\sigmaL}\not\ll 1$, and they cannot occur at the same time. Let us work out the former case first. When $\xi_{\sigmaR}\not\ll1$, the third integral in \Eq{A.2} is negligible and the first one is almost calculated in \Eq{steepest}, so the second one is focused on. Recalling that the second derivative of $f$ is taken into account, instead of \Eq{firstderapp1}, 
\begin{align}
    &\int_{\sigmaR}^{\infty} \dd\sigma e^{f(\sigma)/\epsilon} g(\sigma) 
    = \int_{\sigmaR}^{\infty} \dd \sigma \exp\left[\frac{f_\mathrm{R}}{\epsilon} + \frac{f'_\mathrm{R}}{\epsilon} (\sigma-\sigmaR) + \frac{f''_\mathrm{R}}{2!\epsilon} (\sigma-\sigmaR)^2 + \cdots\right] [g_\mathrm{R} + g'_\mathrm{R} (\sigma-\sigmaR) + \cdots] \notag\\
    &= e^{\xi_{\sigmaR}^{-1}/2 + f_\mathrm{R}/\epsilon} \sqrt{\frac{2 \epsilon}{-f_\mathrm{R}''}} \int_0^{\infty} \dd z \exp\left[ - \left( z + \frac{\xi_{\sigmaR}^{-1/2}}{2} \right)^2  \right] \left(1+ \frac{f_\mathrm{R}'''}{3!\epsilon} \sqrt{\frac{2 \epsilon}{-f''}}^3 z^3 +\cdots\right) \left(g + g' \sqrt{\frac{2 \epsilon}{-f''}} z + \cdots\right) \notag\\
    &= e^{f_\mathrm{R}/\epsilon} \sqrt{\frac{2 \epsilon}{-f''_\mathrm{R}}} \left( \frac{g_\mathrm{R}\sqrt{\pi}}{2}  e^{z_{\sigmaR}^2}\erfc(z_{\sigmaR}) + \sqrt{\frac{2\epsilon}{-f''_\mathrm{R}}}\left\{\frac{g'_\mathrm{R}}{2}\left[1 - z_{\sigmaR}\sqrt{\pi}e^{z_{\sigmaR}^2}\erfc(z_{\sigmaR}) \right] \right.\right.\notag\\
    &\hspace{20mm} \left.\left. + \frac{g_\mathrm{R}f'''_\mathrm{R}}{3!(-f''_\mathrm{R})}\left[(1+z_{\sigmaR}^2) - \frac{z_{\sigmaR}}{2}(3+2z_{\sigmaR})\sqrt{\pi}e^{z_{\sigmaR}^2}\erfc(z_{\sigmaR}) \right] \right\} + \mathcal{O}(\epsilon) \right) , \label{secondderapp1}
\end{align}
 where $z_{\sigmaR} \equiv \xi_{\sigmaR}^{-1/2}/2$. Since $f'(\sigmaR)$ and $ f''(\sigmaR)$ are not very small simultaneously (more precisely, when $\xi_{\sigmaR} \not\ll 1 $, $f''(\sigmaR)< 0$ in the limit $\epsilon\ll1$), it is sufficient to use \Eq{secondderapp1} as a complementary formula, which cannot be applied to \Eq{firstderapp1} used when $\xi_{\sigmaR} \ll 1$. Let us note that as $f'(\sigmaR)$ is far from 0, $z_{\sigmaR}$ is much lager than 1. Consequently, it is easy to check that \Eq{secondderapp1} reduces to \Eq{firstderapp1} by expanding the complementary error function asymptotically, although this expansion makes \eqref{secondderapp1} no longer meaningful, since $\epsilon$ in $z_{\sigmaR}$ have to be considered. Combining \Eqs{steepest} and~\eqref{secondderapp1} up to $e^{f/\epsilon}\mathcal{O}(\epsilon)$,
 \begin{align}
     &e^{f_0/\epsilon} \sqrt{\frac{2\pi \epsilon}{- f''_0}} g_0 - e^{f_\mathrm{R}/\epsilon} \sqrt{\frac{2 \epsilon}{-f''_\mathrm{R}}} \left( \frac{g\sqrt{\pi}}{2}  e^{z_{\sigmaR}^2}\erfc(z_{\sigmaR}) + \sqrt{\frac{2\epsilon}{-f''_\mathrm{R}}}\left\{\frac{g'_\mathrm{R}}{2}\left[1 - z_{\sigmaR}\sqrt{\pi}e^{z_{\sigmaR}^2}\erfc(z_{\sigmaR}) \right]\right.\right.\notag\\
    &\hspace{30mm} \left.\left. + \frac{g_\mathrm{R}f'''_\mathrm{R}}{3!(-f''_\mathrm{R})}\left[(1+z_{\sigmaR}^2) - \frac{z_{\sigmaR}}{2}(3+2z_{\sigmaR})\sqrt{\pi}e^{z_{\sigmaR}^2}\erfc(z_{\sigmaR}) \right] \right\} \right),
 \end{align}
 which leads to \Eq{NLOneq5}. 

 When $\xi_{\sigmaL}\not\ll1$, the calculation can be performed along similar lines. By replacing $\sigmaR \rightarrow \sigmaL$ and flipping the signature of terms of order $\mathcal{O}(\epsilon)$ in \Eq{secondderapp1}, the result is obtained, leading to \Eq{NLOneq3}.

\subsection*{$(\mathrm{ii})\ \sigma_{0} < \sigmaL$}
The integral \eqref{NLOoriginal} is rewritten as 
\begin{align}
    \int_{\sigmaL}^{\sigmaR} \dd\sigma e^{f(\sigma)/\epsilon} g(\sigma) = \left(\int_{\sigmaL}^{\infty} - \int_{\sigmaR}^{\infty} \right) \dd\sigma e^{f(\sigma)/\epsilon} g(\sigma) . \label{A.9}
\end{align}
In the limit $\epsilon\ll1$, the second integral is negligible compared to the first one. When $\xi_{\sigmaL}\ll 1$, these integrals are the same form as \Eq{firstderapp1}. Therefore, by replacing $\sigmaL$ with $\sigmaR$ in \Eq{firstderapp1}, one obtains \Eq{NLOneq1}. When $\xi_{\sigmaL}\not\ll 1$, these integrals have the exact same form as \Eq{secondderapp1}, and then by replacing $\sigmaL$ with $\sigmaR$ in \Eq{firstderapp1}, one obtains \Eq{NLOneq2}. 

\subsection*{$(\mathrm{iii})\ \sigma_{0} > \sigmaR$}
The integral \eqref{NLOoriginal} is rewritten as 
\begin{align}
    \int_{\sigmaL}^{\sigmaR} \dd\sigma e^{f(\sigma)/\epsilon} g(\sigma) = \left(\int^{\sigmaR}_{-\infty} - \int^{\sigmaL}_{-\infty} \right) \dd\sigma e^{f(\sigma)/\epsilon} g(\sigma) . \label{A.10}
\end{align}
In the limit $\epsilon\ll1$, the second integral is negligible compared to the first one. Basically, this case is similar to the previous case. Proceeding along the same lines, in agreement with the value of $\xi_{\sigmaR}$, \Eqs{NLOneq6} and~\eqref{NLOneq7} are obtained.

\section{Analytical expressions for the distribution functions}
\label{app:analytical:appr}
This section presents the full expressions of the probability density functions (PDF) for the total number of \efolds;  $P(\mathcal{N}|\phi_*,\sigma_*)$ and for the joint probability of total and inflationary \efolds; $P(\mathcal{N},\mathcal{N}_\mathrm{inf})$, discussed in \Sec{Sec:4}. {These formulas have been used to produce \Figs{Fig:noneq} and \ref{Fig:eqjoint}.}

\subsection{$P(\mathcal{N}|\phi_*,\sigma_*)$}\label{Sec:C}
A straightforward calculation gives rise to the expression of the contribution from scenario 4, $P^{(4)}(\mathcal{N}|\phi_*,\sigma_*)$;
\begin{align}
    P^{(4)}(\mathcal{N}|\phi_*,\sigma_*) &= \frac{1}{2}\left\{\mathrm{erf}\left[\frac{\Braket{\sigma(\mathcal{N} - N^{(4)}_{\mathrm{post}})} + \sigmaL}{\sqrt{2\Braket{\delta\sigma^2(\mathcal{N} - N^{(4)}_{\mathrm{post}})}}}\right] - \mathrm{erf}\left[\frac{\Braket{\sigma(\mathcal{N} - N^{(4)}_{\mathrm{post}})} - \sigmaL}{\sqrt{2\Braket{\delta\sigma^2(\mathcal{N} - N^{(4)}_{\mathrm{post}})}}}\right]\right\}\notag\\
    &\quad \times\frac{1}{\sqrt{2\pi\Braket{\delta{\mathcal{N}_\mathrm{inf}}^2}}}\exp\left[\frac{1}{2\Braket{\delta{\mathcal{N}_\mathrm{inf}}^2}}(\mathcal{N} - N_{\mathrm{post}}^{(4)} - \Braket{\mathcal{N}_{\mathrm{inf}}})^2\right].\label{NLOcase4noneq}
\end{align}
In contrast, the contribution from scenario 5 is not calculable without approximations. However, when $\Braket{\delta{\mathcal{N}_\mathrm{inf}}^2}\ll1$, the steepest-descent method discussed in \App{Sec:A} can be used. As a result, the expression of the contribution from scenario 5, $P^{(5)}(\mathcal{N}|\phi_*,\sigma_*)$ is given as follows.\footnote{Since $\sigma_0$, where $f(\sigma_{\mathrm{end}})$ reaches a local maximum, depends on $\mathcal{N}$, $P^{(5)}(\mathcal{N}|\phi_*,\sigma_*)$ has to be calculated along the several cases listed in \App{Sec:A}. For brevity, the notations for $\sigma,\epsilon,f,\xi,z$ and $N_{\mathrm{post,R}}^{(5)}\equiv N_{\mathrm{post}}^{(5)} (\sigmaR)$ are used in the following in the same way as \App{Sec:A}.}

\subsection*{$(\mathrm{i})\ N_{\mathrm{post}}^{(4)} < \mathcal{N} -\Braket{\mathcal{N}_{\mathrm{inf}}} <  N_{\mathrm{post,R}}^{(5)}$}
\textbf{$\bullet\ \xi_{\sigmaL}(\mathcal{N})\not\ll1$}
    \begin{align}
   &\frac{3}{\sqrt{8 \pi}} \frac{\sigmaL}{\sqrt{\Braket{\delta\sigma^2(\Braket{\mathcal {N}_{\mathrm{inf}}})}}}
    \exp\left[ \frac{3}{2}(\mathcal{N} - N_{\mathrm{post}}^{(4)} - \Braket{\mathcal {N}_{\mathrm{inf}}}) \right] 
    \exp\left\{-\frac{\left[\sigmaL e^{\frac{3}{2}(\mathcal{N} - N_{\mathrm{post}}^{(4)} - \Braket{\mathcal {N}_{\mathrm{inf}}})} - \Braket{\sigma(\Braket{\mathcal {N}_{\mathrm{inf}}})}\right]^2}{2\Braket{\delta\sigma^2(\Braket{\mathcal {N}_{\mathrm{inf}}})}}\right\}\notag\\
    &- \frac{1}{2\pi} \frac{\exp\left\{-\frac{\left[\sigmaL - \Braket{\sigma\left(\mathcal{N} - N_{\mathrm{post}}^{(4)}\right)}\right]^2}{2\Braket{\delta\sigma^2\left(\mathcal{N} - N_{\mathrm{post}}^{(4)}\right)}}\right\}  }{\sqrt{ (-f''_\mathrm{L})\Braket{\delta\sigma^2\left(\mathcal{N} - N_{\mathrm{post}}^{(4)}\right)}}}   \exp\left[- \frac{\left(\mathcal{N} - N_{\mathrm{post}}^{(4)} - \Braket{\mathcal{N}_{\mathrm{inf}}}\right)^2}{2\epsilon}\right] \notag\\
    &\times   \left( \frac{\sqrt{\pi}}{2}  e^{z_{\sigmaL}^2}\erfc(z_{\sigmaL}) \right.- \sqrt{\frac{\epsilon}{-f''_\mathrm{L}}}\left\{\frac{-\sigmaL+ \Braket{\sigma\left(\mathcal{N} - N_{\mathrm{post}}^{(4)}\right)} }{ \Braket{\delta\sigma^2\left(\mathcal{N} - N_{\mathrm{post}}^{(4)}\right)} }\left[1 - z_{\sigmaL}\sqrt{\pi}e^{z_{\sigmaL}^2}\erfc(z_{\sigmaL}) \right] \right.\notag\\
    &\quad + \left.\left.\frac{f'''_\mathrm{L}}{6(-f''_\mathrm{L})}\left[(1+z_{\sigmaL}^2) - \frac{z_{\sigmaL}}{2}(3+2z_{\sigmaL})\sqrt{\pi}e^{z_{\sigmaL}^2}\erfc(z_{\sigmaL}) \right] \right\}\right)\notag\\
    & + \left(\mathrm{terms\ where}\Braket{\sigma} \mathrm{replaced\ with} - \Braket{\sigma} \right),\label{NLOneq3}
    \end{align}
\textbf{$\bullet\ \xi_{\sigmaL}(\mathcal{N}),\xi_{\sigmaR}(\mathcal{N})\ll 1$}
    \begin{align}
    &\frac{3}{\sqrt{8 \pi}}  \frac{\sigmaL}{\sqrt{\Braket{\delta\sigma^2(\Braket{\mathcal {N}_{\mathrm{inf}}})}}}
    \exp\left[ \frac{3}{2}(\mathcal{N} - N_{\mathrm{post}}^{(4)} - \Braket{\mathcal {N}_{\mathrm{inf}}}) \right] 
    \exp\left\{-\frac{\left[\sigmaL e^{\frac{3}{2}(\mathcal{N} - N_{\mathrm{post}}^{(4)} - \Braket{\mathcal {N}_{\mathrm{inf}}})} - \Braket{\sigma(\Braket{\mathcal {N}_{\mathrm{inf}}})}\right]^2}{2\Braket{\delta\sigma^2(\Braket{\mathcal {N}_{\mathrm{inf}}})}}\right\}\notag\\
    &- \frac{3\sqrt{\epsilon} }{4\pi } \frac{ \sigmaL }{\sqrt{ \Braket{\delta\sigma^2\left(\mathcal{N} - N_{\mathrm{post}}^{(4)}\right)}}}    \frac{e^{\left\{-\frac{\left[\sigmaL - \Braket{\sigma\left(\mathcal{N} - N_{\mathrm{post}}^{(4)}\right)}\right]^2}{2\Braket{\delta\sigma^2\left(\mathcal{N} - N_{\mathrm{post}}^{(4)}\right)}}\right\}}}{\mathcal{N} -  \Braket{\mathcal{N}_{\mathrm{inf}}} -N_{\mathrm{post}}^{(4)}  }  \exp\left[- \frac{\left(\mathcal{N} - N_{\mathrm{post}}^{(4)} - \Braket{\mathcal{N}_{\mathrm{inf}}}\right)^2}{2\epsilon}\right] \notag\\
    &-\frac{3\sqrt{\epsilon} }{4\pi } \frac{\sigmaR }{\sqrt{ \Braket{\delta\sigma^2\left(\mathcal{N} - N_{\mathrm{post,R}}^{(5)}\right)}}}   \frac{e^{\left\{-\frac{\left[\sigmaR - \Braket{\sigma\left(\mathcal{N} - N_{\mathrm{post,R}}^{(5)}\right)}\right]^2}{2\Braket{\delta\sigma^2\left(\mathcal{N} - N_{\mathrm{post,R}}^{(5)}\right)}}\right\}}}{N_{\mathrm{post,R}}^{(5)}- \mathcal{N} +  \Braket{\mathcal{N}_{\mathrm{inf}}} } \exp\left[- \frac{\left(\mathcal{N} - N_{\mathrm{post,R}}^{(5)}- \Braket{\mathcal{N}_{\mathrm{inf}}}\right)^2}{2\epsilon}\right] \notag\\
    & + \left(\mathrm{terms\ where}\Braket{\sigma} \mathrm{replaced\ with} - \Braket{\sigma} \right),\label{NLOneq4}
    \end{align}
\textbf{$\bullet\ \xi_{\sigmaR}(\mathcal{N})\not\ll 1$}
    \begin{align}
    &\frac{3}{\sqrt{8 \pi}} \frac{\sigmaL}{\sqrt{\Braket{\delta\sigma^2(\Braket{\mathcal {N}_{\mathrm{inf}}})}}}
    \exp\left[ \frac{3}{2}(\mathcal{N} - N_{\mathrm{post}}^{(4)} - \Braket{\mathcal {N}_{\mathrm{inf}}}) \right] 
    \exp\left\{-\frac{\left[\sigmaL e^{\frac{3}{2}(\mathcal{N} - N_{\mathrm{post}}^{(4)} - \Braket{\mathcal {N}_{\mathrm{inf}}})} - \Braket{\sigma(\Braket{\mathcal {N}_{\mathrm{inf}}})}\right]^2}{2\Braket{\delta\sigma^2(\Braket{\mathcal {N}_{\mathrm{inf}}})}}\right\}\notag\\
    &- \frac{1}{2\pi }\frac{\exp\left\{-\frac{\left[\sigmaR - \Braket{\sigma\left(\mathcal{N} - N_{\mathrm{post,R}}^{(5)}\right)}\right]^2}{2\Braket{\delta\sigma^2\left(\mathcal{N} - N_{\mathrm{post,R}}^{(5)}\right)}}\right\} }{\sqrt{(-f''_\mathrm{R})\Braket{\delta\sigma^2\left(\mathcal{N} - N_{\mathrm{post,R}}^{(5)}\right)}}}   \exp\left[- \frac{\left(\mathcal{N} - N_{\mathrm{post,R}}^{(5)}- \Braket{\mathcal{N}_{\mathrm{inf}}}\right)^2}{2\epsilon}\right] \notag\\
    &\times\left( \frac{\sqrt{\pi}}{2}  e^{z_{\sigmaR}^2} \erfc(z_{\sigmaR}) \right.+ \sqrt{\frac{\epsilon}{-f''_\mathrm{R}}}\left\{\frac{-\sigmaR + \Braket{\sigma\left(\mathcal{N} - N_{\mathrm{post,R}}^{(5)}\right)} }{ \Braket{\delta\sigma^2\left(\mathcal{N} - N_{\mathrm{post,R}}^{(5)}\right)} }\left[1 - z_{\sigmaR}\sqrt{\pi}e^{z_{\sigmaR}^2} \erfc(z_{\sigmaR}) \right]\right.\notag\\
    &\quad + \left.\left.\frac{f'''_\mathrm{R}}{3!(-f''_\mathrm{R})}\left[(1+z_{\sigmaR}^2) - \frac{z_{\sigmaR}}{2}(3+2z_{\sigmaR})\sqrt{\pi}e^{z_{\sigmaR}^2} \erfc(z_{\sigmaR}) \right] \right\}\right)\notag\\
    & + \left(\mathrm{terms\ where}\Braket{\sigma} \mathrm{replaced\ with} - \Braket{\sigma} \right),\label{NLOneq5}
    \end{align}

\subsection*{$(\mathrm{ii})\ \mathcal{N} -\Braket{\mathcal{N}_{\mathrm{inf}}} < N_{\mathrm{post}}^{(4)}$}
\textbf{$\bullet\ \xi_{\sigmaL}(\mathcal{N})\ll 1$}
    \begin{align}
    &\frac{3\sqrt{\epsilon}}{4\pi} \frac{\sigmaL}{\sqrt{\Braket{\delta\sigma^2\left(\mathcal{N} - N_{\mathrm{post}}^{(4)}\right)}}}   \frac{\exp\left\{-\frac{\left[\sigmaL - \Braket{\sigma\left(\mathcal{N} - N_{\mathrm{post}}^{(4)}\right)}\right]^2}{2\Braket{\delta\sigma^2\left(\mathcal{N} - N_{\mathrm{post}}^{(4)}\right)}}\right\}  }{N_{\mathrm{post}}^{(4)} - \mathcal{N} +\Braket{\mathcal{N}_{\mathrm{inf}}} } \exp\left[- \frac{\left(\mathcal{N} - N_{\mathrm{post}}^{(4)} - \Braket{\mathcal{N}_{\mathrm{inf}}}\right)^2}{2\epsilon}\right]\notag\\
    & + \left(\mathrm{a\ term\ where}\Braket{\sigma} \mathrm{replaced\ with} - \Braket{\sigma} \right),\label{NLOneq1}
    \end{align}
\textbf{$\bullet\ \xi_{\sigmaL}(\mathcal{N})\not\ll 1$}
    \begin{align}
    & \frac{1 }{2\pi } \frac{ \exp\left\{-\frac{\left[\sigmaL - \Braket{\sigma\left(\mathcal{N} - N_{\mathrm{post}}^{(4)}\right)}\right]^2}{2\Braket{\delta\sigma^2\left(\mathcal{N} - N_{\mathrm{post}}^{(4)}\right)}}\right\} }{\sqrt{(-f''_\mathrm{L})) \Braket{\delta\sigma^2\left(\mathcal{N} - N_{\mathrm{post}}^{(4)}\right)}}} \exp\left[- \frac{\left(\mathcal{N} - N_{\mathrm{post}}^{(4)} - \Braket{\mathcal{N}_{\mathrm{inf}}}\right)^2}{2\epsilon}\right] \notag\\
    &\times  \left( \frac{\sqrt{\pi}}{2}  e^{z_{\sigmaL}^2}\erfc(z_{\sigmaL}) \right.  + \sqrt{\frac{\epsilon}{-f''_\mathrm{L}}}\left\{\frac{-\sigmaL+ \Braket{\sigma\left(\mathcal{N} - N_{\mathrm{post}}^{(4)}\right)} }{ \Braket{\delta\sigma^2\left(\mathcal{N} - N_{\mathrm{post}}^{(4)}\right)} }\left[1 - z_{\sigmaL}\sqrt{\pi}e^{z_{\sigmaL}^2}\erfc(z_{\sigmaL}) \right] \right.\notag\\
    &\quad + \left.\left.\frac{f'''_\mathrm{L}}{6(-f''_\mathrm{L})}\left[(1+z_{\sigmaL}^2) - \frac{z_{\sigmaL}}{2}(3+2z_{\sigmaL})\sqrt{\pi}e^{z_{\sigmaL}^2}\erfc(z_{\sigmaL}) \right] \right\}\right)\notag\\
    & + \left(\mathrm{a\ term\ where}\Braket{\sigma} \mathrm{replaced\ with} - \Braket{\sigma} \right),\label{NLOneq2}
    \end{align}

\subsection*{$(\mathrm{iii})\  N_{\mathrm{post,R}}^{(5)}< \mathcal{N} 
    - \Braket{\mathcal{N}_{\mathrm{inf}}}$}
\textbf{$\bullet\ \xi_{\sigmaR}(\mathcal{N})\not\ll1$}
    \begin{align}
    &\frac{1 }{2\pi } \frac{\exp\left\{-\frac{\left[\sigmaR - \Braket{\sigma\left(\mathcal{N} - N_{\mathrm{post,R}}^{(5)}\right)}\right]^2}{2\Braket{\delta\sigma^2\left(\mathcal{N} - N_{\mathrm{post,R}}^{(5)}\right)}}\right\} }{\sqrt{(-f''_\mathrm{R}) \Braket{\delta\sigma^2\left(\mathcal{N} - N_{\mathrm{post,R}}^{(5)}\right)}}}  \exp\left[- \frac{\left(\mathcal{N} - N_{\mathrm{post,R}}^{(5)}- \Braket{\mathcal{N}_{\mathrm{inf}}}\right)^2}{2\epsilon}\right] \notag\\
    &\times  \left( \frac{\sqrt{\pi}}{2}  e^{z_{\sigmaR}^2} \erfc(z_{\sigmaR}) \right. - \sqrt{\frac{\epsilon}{-f''_\mathrm{R}}}\left\{\frac{-\sigmaR + \Braket{\sigma\left(\mathcal{N} - N_{\mathrm{post,R}}^{(5)}\right)} }{ \Braket{\delta\sigma^2\left(\mathcal{N} - N_{\mathrm{post,R}}^{(5)}\right)} }\left[1 - z_{\sigmaR}\sqrt{\pi}e^{z_{\sigmaR}^2} \erfc(z_{\sigmaR}) \right] \right.\notag\\
    & \quad + \left.\left.\frac{f'''_\mathrm{R}}{3!(-f''_\mathrm{R})}\left[(1+z_{\sigmaR}^2) - \frac{z_{\sigmaR}}{2}(3+2z_{\sigmaR})\sqrt{\pi}e^{z_{\sigmaR}^2} \erfc(z_{\sigmaR}) \right] \right\}\right)\notag\\
    & + \left(\mathrm{a\ term\ where}\Braket{\sigma} \mathrm{replaced\ with} - \Braket{\sigma} \right),\label{NLOneq6}
    \end{align}
\textbf{$\bullet\ \xi_{\sigmaR}(\mathcal{N})\ll 1$}
    \begin{align}
    & \frac{3 \sqrt{\epsilon}}{4\pi} \frac{\sigmaR}{\sqrt{ \Braket{\delta\sigma^2\left(\mathcal{N} - N_{\mathrm{post,R}}^{(5)}\right)}}} \frac{\exp\left\{-\frac{\left[\sigmaR - \Braket{\sigma\left(\mathcal{N} - N_{\mathrm{post,R}}^{(5)}\right)}\right]^2}{2\Braket{\delta\sigma^2\left(\mathcal{N} - N_{\mathrm{post,R}}^{(5)}\right)}}\right\} }{\mathcal{N} - \Braket{\mathcal{N}_{\mathrm{inf}}} - N_{\mathrm{post,R}}^{(5)}}  \exp\left[- \frac{\left(\mathcal{N} - N_{\mathrm{post,R}}^{(5)}- \Braket{\mathcal{N}_{\mathrm{inf}}}\right)^2}{2\epsilon}\right] \notag\\
    & + \left(\mathrm{a\ term\ where}\Braket{\sigma} \mathrm{replaced\ with} - \Braket{\sigma} \right).\label{NLOneq7}
    \end{align}

\subsection{$P(\mathcal{N},\mathcal{N}_\mathrm{inf})$ }\label{Sec:B2}
Without mathematical approximations, \Eq{joint probability efold} can be calculated as follows, 
\begin{align}
    P^{(4)}(\mathcal{N},\mathcal{N}_\mathrm{inf})&= \frac{1}{2}\left\{\mathrm{erf}\left[\frac{\Braket{\sigma(\mathcal{N} - N^{(4)}_{\mathrm{post}})} + \sigmaL}{\sqrt{2\Braket{\sigma^2(\mathcal{N} - N^{(4)}_{\mathrm{post}})}}}\right] - \mathrm{erf}\left[\frac{\Braket{\sigma(\mathcal{N} - N^{(4)}_{\mathrm{post}})} - \sigmaL}{\sqrt{2\Braket{\sigma^2(\mathcal{N} - N^{(4)}_{\mathrm{post}})}}}\right]\right\}\notag\\
    &\quad\quad\times P_{\mathrm{FPT}}(\mathcal{N}_\mathrm{inf};\phi_*) \delta\left(\mathcal{N}_\mathrm{inf} + N_\mathrm{post}^{(4)} -\mathcal{N} \right)\label{njointcase4}
\end{align}
and
\begin{align}
    P^{(5)}(\mathcal{N},\mathcal{N}_\mathrm{inf}) &=    P_{\mathrm{FPT}}(\mathcal{N}_\mathrm{inf};\phi_*)\frac{3}{\sqrt{8\pi}}\frac{\sigmaL}{\Braket{\delta\sigma^2(\mathcal{N}_\mathrm{inf})}} \exp\left[{\frac{3}{2}(\mathcal{N}_\mathrm{inf} + N_{\mathrm{post}}^{(4)} - \mathcal {N})}\right]\notag\\
    &\quad\quad\quad\quad\quad\quad\quad \quad\times \exp\left\{-\frac{  \left[\sigmaL e^{\frac{3}{2}(\mathcal{N}_\mathrm{inf} + N_\mathrm{post}^{(4)} - \mathcal {N})} - \Braket{\sigma(\mathcal {N}_\mathrm{inf})}\right]^2  }{2\Braket{\delta\sigma^2(\mathcal {N}_\mathrm{inf})}}\right\} \notag\\
    &\quad+ \left(\mathrm{a\ term\ where}\Braket{\sigma} \mathrm{replaced\ with} - \Braket{\sigma} \right).\label{njointcase5}
\end{align}

\section{Comparison with cosmological perturbation theory}

{In this section, we check that our results reduce to cosmological perturbation theory in the appropriate limit.}
In the context of cosmological perturbation theory, the curvature perturbations on the curvaton decay surface are often characterised by a local-type non-Gaussianity in the limit where the inflaton contributions are negligible. Given $\zeta = \zeta_\mathrm{G} + \frac{3}{5}f_{\mathrm{NL}}\left(\zeta_\mathrm{G}^2 - \overline{\zeta_\mathrm{G}^2}\right)$ where $\zeta_\mathrm{G}$ obeys the Gaussian statics and a quantity with the overline $\overline{\cdot}$ represents a spatial average (or a quantum expectation value), the PDF of local-type non-Gaussian distribution up to quadratic order; $P_2(\zeta)$ is obtained by substituting $\zeta_\mathrm{G} =\zeta - \frac{3}{5}f_{\mathrm{NL}}\left(\zeta^2 - \overline{\zeta_\mathrm{G}^2}\right) + \mathcal{O}(\zeta^3)$ into the Gaussian distribution and multiplying this by a Jacobian coming from the change of variable leads to\footnote{For simplicity, one branch in $\zeta_\mathrm{G}(\zeta)$ is considered. If taking into account another branch, another term must be added to \Eq{local2}.\label{footnote:branch}}
\begin{align}
    P_2(\zeta) = \frac{1}{\sqrt{2\pi \overline{\zeta_\mathrm{G}^2}}} \exp\left[-\frac{9}{5}f_{\mathrm{NL}}\zeta-\frac{1}{2\overline{\zeta_\mathrm{G}^2}}\left(\zeta^2 -\frac{6}{5}f_{\mathrm{NL}}\zeta^3\right)\right].\label{local2}
\end{align}
If considering the quadratic potential for the curvaton, {the evolution from Hubble exit when $\sigma_{*} = \overline \sigma_{*} + \delta \sigma_{*}$ to the phase of curvaton oscillation is linear,}
hence $\zeta_{\mathrm{G}}$ and $f_\mathrm{NL}$ are expressed as
\begin{align}
    \zeta_{\mathrm{G}} = \frac{2}{3}r_{\mathrm{dec}} \frac{\delta_1\sigma}{\overline\sigma} \label{zetacurvaton}, 
\end{align}
and
\begin{align}
    f_{\mathrm{NL}} = \frac{5}{4r_\mathrm{dec}} - \frac{5}{3} - \frac{5r_\mathrm{dec}}{6}\label{fnl},
\end{align}
where $\overline \sigma$, $\delta_1 \sigma$ are the background field and the linear perturbation of $\sigma$ respectively \cite{Sasaki:2006kq}. 

{In the stochastic $\delta N$ formalism, the non-linear dynamics of both the inflaton and the curvaton fields at large scales are accounted for, and non-linear effects are crucial to determine the shape of the tail of the curvature perturbation. However, close to the peak of the distribution, they are expected to play a more minor role if fluctuations remain small. Our goal is to check that our results reduce to cosmological perturbation theory in that limit.}

A comparison can be made between the results under the stochastic formalism obtained in this paper and those predicted by the cosmological perturbation theory. The stochastic formalism should be reduced to the cosmological perturbation theory in the limit where the stochastic diffusion term is negligible, that is, fluctuations of the curvaton is very small comparing the background field. Therefore, let us consider $\varepsilon \equiv \Braket{\delta\sigma^2(\Braket{\mathcal {N}_{\mathrm{inf}}})}/ \Braket{\sigma(\Braket{\mathcal {N}_{\mathrm{inf}}})}^2  $ to be very small and expand quantities in $\varepsilon$. In this limit, there shall be no non-linear evolution in the case of a quadratic potential, which is same as the result of the cosmological perturbation theory. Let us also take the limit of $\Braket{\delta{\mathcal{N}_\mathrm{inf}}^2}\rightarrow 0$ so that perturbations of the inflaton vanish. Considering the situation of the usual curvaton scenario where the energy of the curvaton is dominant when it decays, \ie the ``scenario 5", only the first term in \Eqs{NLOneq3}-\eqref{NLOneq5} have to be taken into account in the PDF of $\mathcal{N}$,
\begin{align}
    P(\mathcal{N})  \notag
    = &\frac{3}{\sqrt{8 \pi}} \frac{\sigmaL}{\sqrt{\Braket{\delta\sigma^2(\Braket{\mathcal {N}_{\mathrm{inf}}})}}}
    \exp\left[ \frac{3}{2}(\mathcal{N} - N_{\mathrm{post}}^{(4)} - \Braket{\mathcal {N}_{\mathrm{inf}}}) \right] 
    \\ & \times
    \exp\left\{-\frac{\left[\sigmaL e^{\frac{3}{2}(\mathcal{N} - N_{\mathrm{post}}^{(4)} - \Braket{\mathcal {N}_{\mathrm{inf}}})} - \Braket{\sigma(\Braket{\mathcal {N}_{\mathrm{inf}}})}\right]^2}{2\Braket{\delta\sigma^2(\Braket{\mathcal {N}_{\mathrm{inf}}})}}\right\} \notag\\
    &=  \frac{1}{\sqrt{2\pi \left(\frac{4}{9}\varepsilon\right)}} e^{\frac{3}{2}\left[\mathcal{N} - N_{\mathrm{post}}^{(4)} - \Braket{\mathcal N}_{\mathrm{inf}} -\frac{2}{3}\ln\left(\frac{\Braket{\sigma}}{\sigmaL}\right) \right]} \exp\left(-\frac{\left\{\frac{2}{3}e^{\frac{3}{2}\left[\mathcal{N} - N_{\mathrm{post}}^{(4)} - \Braket{\mathcal N}_{\mathrm{inf}} -\frac{2}{3}\ln\left(\frac{\Braket{\sigma}}{\sigmaL}\right) \right]} - \frac{2}{3}\right\}^2}{2\left(\frac{4}{9}\varepsilon\right)}\right). \label{stochasticcurvaton}
\end{align}
Here, $\sigmaL<\Braket{\sigma}<\sigmaR$ is considered. Let us notice that in the contribution over $-\sigmaR<\Braket{\sigma}<-\sigmaL$, a term where $\Braket{\sigma}$ is replaced with $-\braket{\sigma}$ in \Eq{NLOneq3} is neglected for simplicity, which corresponds to the other branch mentioned in the footnote \ref{footnote:branch}.  This can be written as a local type non-Gaussian distribution as follows.

Firstly, $\braket{\mathcal{N}} = \int \dd \mathcal{N} P(\mathcal{N})\mathcal{N}$ can be expanded in $\varepsilon$ by making use of the steepest-decent method \eqref{steepest} and the result is
\begin{align}
    \braket{\mathcal{N}}
    &= \Braket{\mathcal{N}_{\mathrm{inf}}} + N_{\mathrm{post}}^{(4)} + \frac{2}{3}\ln\left(\frac{\Braket{\sigma}}{\sigmaL}  \right) - \frac{1}{3} \varepsilon + \mathcal{O}(\varepsilon^{2}). 
\end{align}
Recalling that curvature perturbations can be expressed as $\zeta=\delta\mathcal{N}=\mathcal{N}- \Braket{\mathcal{N}}$, its distribution function is given by
\begin{align}
    P(\zeta) = \frac{1 - \frac{1}{2}\varepsilon + \mathcal{O}(\varepsilon^{2})}{\sqrt{2\pi \left(\frac{4}{9}\varepsilon\right)}} \exp\left(\frac{3}{2}\zeta-\frac{1}{2\left(\frac{4}{9}\varepsilon\right)}\left\{\frac{2}{3}\left[1 - \frac{1}{2}\varepsilon + \mathcal{O}(\varepsilon^{2})\right]e^{\frac{3}{2}\zeta} - \frac{2}{3}\right\}^2\right). \label{D.3}
\end{align}
When $\varepsilon\ll 1$, the second term of the argument of the exponential in \Eq{D.3} determines the behaviour of $P(\zeta)$. Hence it is found that $P(\zeta) $ takes most support within $|\zeta|\lesssim\varepsilon^{1/2}$ and that $\zeta$ can be regarded as $\mathcal{O}(\varepsilon^{1/2})$. Therefore, considering terms in \Eq{D.3} up to $\mathcal{O}(1)$, $P(\zeta)$ reads 
\begin{align}
    P(\zeta) = \frac{1}{\sqrt{2\pi \left(\frac{4}{9}\varepsilon\right)}} \exp\left[\frac{9}{4}\zeta-\frac{1}{2\left(\frac{4}{9}\varepsilon\right)}\left(\zeta^2 + \frac{3}{2}\zeta^3\right)\right]\label{D.4},
\end{align}
which is a local-type non-Gaussian distribution \eqref{local2} with these identification; $\overline{\zeta_\mathrm{G}^2}= \frac{4}{9}\varepsilon$ and $f_{\mathrm{NL}}=-\frac{5}{4}$. These values exactly correspond to \Eqs{zetacurvaton} and~\eqref{fnl} in the scenario 5 of the curvaton, $r_\mathrm{dec}\sim 1$. 

Let us finally emphasize again that \Eq{stochasticcurvaton} includes the non-linear evolution coming from the stochastic noise. If one takes the limit of no stochastic diffusion and the curvature perturbations can be treated perturbatively, it can be explicitly demonstrated that \Eq{stochasticcurvaton} is reduced to a local-type non-Gaussian distribution up to the order we want. This is shown above for quadratic order but can be extended to cubic order; $\zeta = \zeta_\mathrm{G} + \frac{3}{5}f_{\mathrm{NL}}(\zeta_\mathrm{G}^2 - \overline{\zeta_\mathrm{G}^2}) + \frac{9}{25}g_{\mathrm{NL}}\zeta_\mathrm{G}^3$ with $g_{\mathrm{NL}} = {25}/{12}$, which matches the result of \cite{Sasaki:2006kq} in the scenario 5.

\bibliographystyle{JHEP}
\bibliography{v2}
\end{document}